\documentclass{sig-alternate}

\usepackage{amsmath}
\usepackage{amssymb}
\usepackage{balance}  
\usepackage{color}
\usepackage[colorlinks=true,linkcolor=black,citecolor=black,pdfpagelabels=false]{hyperref}
\usepackage{times}
\usepackage[TABBOTCAP]{subfigure}
\usepackage{url}
\usepackage[table]{xcolor} \rowcolors{2}{gray!25}{white}

\newcommand{\TechReportVersion}[1]{#1}\newcommand{\PaperVersion}[1]{}

\definecolor{darkgreen}{rgb}{0,.5,.2}
\definecolor{gray}{rgb}{.5,.5,.5}
\definecolor{darkgray}{rgb}{.3,.3,.3}

{\vspace{5mm}\begin{sloppypar}
  \noindent\hrulefill\, BEGIN \,\hrulefill
  \end{sloppypar}
}%
{\begin{sloppypar}
  \noindent\hrulefill\, END   \,\hrulefill
  \end{sloppypar}\vspace{5mm}
}

\newcommand{\todo}[1]{}

\newcommand{\Todo}[2]{#1}

\newcounter{question}

\newcounter{comment}

\newcommand{\removable}[1]{#1}
\newcommand{\hidden}[1]{}

\newcommand{\removableAltOff}[2]{\removable{#1}}  
\newcommand{\removableAltOn}[2]{#2}   

\newcommand{\definedAs}    
	{=}

\newcommand{\fctsymDom}{\mathrm{dom}} 
\newcommand{\fctDom}[1]{\fctsymDom(#1)}

\newcommand{\tuple}[1]{\langle #1 \rangle} 

\newcommand{\symAllTerms}{\mathcal{T}} 

\newcommand{\symRDFgraph}{G} 

\newcommand{\fctsymVars}{\mathrm{vars}} 
\newcommand{\fctVars}[1]{\fctsymVars(#1)}
\newcommand{\symTP}{tp} 

\newcommand{\symWoD}{W} 

\newcommand{\symReachCrit}{c} 

\newcommand{\cMatch}{\symReachCrit_\mathsf{Match}}
\newcommand{\enumA}{(i)}
\newcommand{\enumB}{(ii)}
\newcommand{\enumC}{(iii)}

\newdef{definition}{Definition}
\newcommand{\definedTerm}[1]{\emph{#1}}

\newtheorem{example}{Example}

\newtheorem{theoremTheorem}{Theorem}

\newtheorem{propositionTheorem}{Proposition}

\newtheorem{lemmaTheorem}{Lemma}

\newtheorem{corollaryTheorem}{Corollary}

\newtheorem{findingTheorem}{Finding}

\newtheorem{factTheorem}{Fact}

\newdef{problem}{Problem}

\newcommand{\Finding}[1]%
	{\textsl{#1}}

\newcommand{\xxxApproachAcronym}[1]{\textsf{#1}}
\newcommand{\xxxApproachAcronymTextSize}[1]{\textsf{\small #1}}
\newcommand{\baseline}{\xxxApproachAcronymTextSize{baseline}}
\newcommand{\BF}{\xxxApproachAcronymTextSize{BFS}} 
\newcommand{\DF}{\xxxApproachAcronymTextSize{DFS}} 
\newcommand{\random}{\xxxApproachAcronymTextSize{random}}
\newcommand{\oracle}{\xxxApproachAcronymTextSize{oracle}}
\newcommand{\indegree}{\xxxApproachAcronymTextSize{indegree}}
\newcommand{\PageRank}{\xxxApproachAcronymTextSize{PageRank}}
\newcommand{\relOne}{\xxxApproachAcronymTextSize{rel1}}
\newcommand{\relTwo}{\xxxApproachAcronymTextSize{rel2}}
\newcommand{\rccOne}{\xxxApproachAcronymTextSize{rcc1}}
\newcommand{\rccTwo}{\xxxApproachAcronymTextSize{rcc2}}
\newcommand{\is}{\xxxApproachAcronymTextSize{IS}}
\newcommand{\isDecr}{\xxxApproachAcronymTextSize{ISdcr}}
\newcommand{\isrelOne}{\xxxApproachAcronymTextSize{isrel1}}
\newcommand{\isrelTwo}{\xxxApproachAcronymTextSize{isrel2}}
\newcommand{\isrccOne}{\xxxApproachAcronymTextSize{isrcc1}}
\newcommand{\isrccTwo}{\xxxApproachAcronymTextSize{isrcc2}}

\newcommand{\DFXX}{\xxxApproachAcronym{DFS}} 
\newcommand{\randomXX}{\xxxApproachAcronym{random}}
\newcommand{\oracleXX}{\xxxApproachAcronym{oracle}}
\newcommand{\indegreeXX}{\xxxApproachAcronym{indegree}}

\newcommand{\relOneXX}{\xxxApproachAcronym{rel1}}
\newcommand{\relTwoXX}{\xxxApproachAcronym{rel2}}
\newcommand{\rccOneXX}{\xxxApproachAcronym{rcc1}}
\newcommand{\rccTwoXX}{\xxxApproachAcronym{rcc2}}
\newcommand{\isXX}{\xxxApproachAcronym{IS}}
\newcommand{\isDecrXX}{\xxxApproachAcronym{ISdcr}}
\newcommand{\isrelOneXX}{\xxxApproachAcronym{isrel1}}
\newcommand{\isrelTwoXX}{\xxxApproachAcronym{isrel2}}
\newcommand{\isrccOneXX}{\xxxApproachAcronym{isrcc1}}
\newcommand{\isrccTwoXX}{\xxxApproachAcronym{isrcc2}}

\newcommand{\policy}[1]{\textsf{\small #1}}

\newcommand{\drop}%
	{\textsc{dr}-op\-er\-a\-tor}

\newcommand{\tpop}%
	{\textsc{tp}-op\-er\-a\-tor}

\newcommand{\tpops}{{\tpop}s}

\newcommand{\mathURI}[1]%
	{\text{\normalfont\small\textsf{#1}}}

\newcommand{\mathVar}[1]%
	{\text{\small #1}}

\begin{document}

\title{%
	Walking without a Map:
%
	Optimizing Response Times
%
	of\\
%
	Traversal-Based Linked Data Queries}

\toappear{} 

\TechReportVersion{
\subtitle{\vspace{-3mm}(Extended Version*)}

\toappear{*This document is an extended version of a paper
	published in the Proceedings of the 15th International Semantic Web Conference~(ISWC~2016)~\cite{Hartig16:ProceedingsVersion}.
In addition to a more detailed discussion of the experimental results presented in the conference version, this extended version provides an in-depth description of our approach to implement tra\-vers\-al-based query execution, and we present a number of additional experiments.%
}
}

\numberofauthors{2}
\author{
\alignauthor
Olaf Hartig\\
       \affaddr{Dept.\ of Computer and Information Science}\\
       \affaddr{Link\"oping University}\\
       \email{olaf.hartig@liu.se}
\alignauthor
M.~Tamer {\"O}zsu\\
       \affaddr{Cheriton School of Computer Science}\\
       \affaddr{University of Waterloo}\\
       \email{tamer.ozsu@uwaterloo.ca}
}

\maketitle

\begin{abstract}
The emergence of Linked Data on the WWW has spawned research interest in an \emph{online} execution of declarative queries over this data. A particularly interesting approach is traversal-based query execution which fetches data by traversing data links and, thus, is able to make use of up-to-date data from initially unknown data sources. The downside of this approach is the
	delay before the query engine completes a query execution.
In this paper, we address this problem by proposing an approach to
	return as many elements of the result set as soon as possible. The basis of this approach is a traversal strategy that aims to fetch
result-relevant data as early as possible.
The challenge for such a strategy is
	that the query engine does not know a priori
which of the data sources that will be discovered during the query execution contain result-relevant data. We introduce
	\PaperVersion{14 }%
	\TechReportVersion{16 }%
different \hidden{heuristics-based }traversal
	approaches and experimentally study their impact on response times.
Our experiments show that some of the approaches can achieve significant improvements over the baseline
	of looking up URIs on a first-come, first-served~basis.
\TechReportVersion{%
	Additionally, we
		\PaperVersion{show }%
		\TechReportVersion{verify the importance of these approaches by showing }%
	that typical query optimization techniques that focus solely on the process of constructing the query result cannot have any significant impact on the response times of traversal-based query executions.
}

\end{abstract}

\section{Introduction} \label{sec:Intro}

In recent years an increasing amount of \emph{structured} data has been published and interlinked on the World Wide Web (WWW) in adherence to the Linked Data principles~\cite{BernersLee06:LinkedData,Bizer09:TheStorySoFar,Mika12:MetadataStatistics}.
The resulting \emph{Web of Linked Data} that is emerging within the WWW can be modeled as a directed
	graph. Each vertex in this graph represents a Web document that contains
data in the form of a set of RDF triples~\cite{Cyganiak14:RDFConcepts}.
These documents can be retrieved by looking up HTTP-scheme based URIs (%
	that is,
URIs that start with http://). A pair of document vertices is connected by a directed edge if any of the RDF triples in the source document contains a URI
	that, when looked up by an HTTP client, enables the client to discover and retrieve
the target document. We call such a connection a \emph{data~link}.

\TechReportVersion{%
Since its introduction, an increasing number of data providers is contributing to the Web of Linked Data~%
	\cite{Bizer09:TheStorySoFar},
including prominent providers such as the BBC~\cite{Kobilarov09:BBC}, the US Library of Congress~\cite{Ford13:LoCLinkedData}, and Renault~\cite{Servant13:RenaultLinkedData}. Available data covers diverse topics such as genes, proteins, medicine, clinical trials, scientific publications, books, reviews, movies, music, radio and television programs, companies, people, statistical and census data, geographic locations,~etc.%
}

The open availability of all this structured and interlinked data presents an exciting opportunity for building applications that
	use
the data
	and its cross-da\-ta\-set connections
in innovative ways.
This possibility has spawned research interest in approaches to enable these applications to query Linked Data
	from multiple sources
\removable{in a declarative manner}~\cite{Arenas12:SIGMODrecord, Harth12:LDMgmtTutorial, Hartig13:Survey}.
A well-un\-der\-stood approach
	to this end
is to populate a centralized repository
	of Linked Data copied from the WWW.
By using such a
	data warehouse approach
it is possible to provide almost instant query results. This capability comes at the cost of setting up and maintaining the centralized repository. Further limitations of this approach are that legal issues may prevent storing a copy of some of the data in the repository, query results may not reflect the most recent status of the copied data, and new data and data sources cannot be exploited.

To address these limitations a number of recent works adopt an alternative view on querying Linked Data: The idea is to conceive the Web of Linked Data itself as a distributed database in which URI lookups are used to access data sources at query execution time~%
	\cite{Hartig09:QueryingTheWebOfLD, Ladwig10:LinkedDataQueryProcessingStrategies, Ladwig11:SIHJoin, Miranker12:Diamond, Umbrich12:LinkedDataQueriesWithReasoning:Article, Umbrich11:DataSummariesForLDQueryProcessingArticle}.
The goal in this context is to enable use cases for which freshness and discovery of results are more important than short execution
	times.
However,
users may want to start receiving elements of the query result
	set
as soon as possible; the query execution process may even be terminated before it is complete if the available result elements already satisfy the user's information need or some us\-er-spec\-i\-fied time limit has been reached. Hence,
	in live querying the Web of Linked Data
it is important to reduce the
	\emph{response times} of query executions, that is,~the times required to find a particular number of result elements\removable{~(as opposed to the overall time required to complete the query execution)}.

A line of research towards this goal adapts typical database techniques such as data summaries and indexes~\cite{Harth10:DataSummariesForLDQueryProcessing, Ladwig10:LinkedDataQueryProcessingStrategies, Umbrich11:DataSummariesForLDQueryProcessingArticle}. While the resulting approaches return query results that are up-to-date w.r.t.~the indexed data sources, and they may
	\removableAltOff{take advantage of}{use}
summary information to optimize query response times, they
	still suffer from some of the 
limitations of the aforementioned warehouse approach:
	Recent updates of remote data sources may not be reflected~in a local summary or index,
and new data sources cannot be exploited by relying
	\removable{only}
on
	such data structures.

An alternative
	approach
that leverages the characteristics of
	the Web of
Linked Data---in particular, the existence of data links---is \emph{tra\-vers\-al-based query execution}~%
\cite{Hartig11:HeuristicForQueryPlanSelection, Hartig09:QueryingTheWebOfLD, Ladwig10:LinkedDataQueryProcessingStrategies, Ladwig11:SIHJoin, Miranker12:Diamond}.
The general idea of this approach is to integrate
	a traversal of data links into the query execution process. Hence, this approach discovers
data and data sources on the fly, and it can be used to start querying right away~(without first having to populate an index or a centralized collection of data).
	However, the response times of tra\-vers\-al-based query executions over the Web of Linked Data may vary significantly as the following example demonstrates.

\begin{example} \label{ex:Motivation}
Consider the following SPARQL query~\cite{Harris13:SPARQL1_1Language} from the FedBench benchmark suite~\cite{Schmidt11:FedBench}
	(which is a benchmark that focuses on
		querying a federation of SPARQL query services).

\vspace{1mm}
\noindent
\begin{scriptsize}\normalfont
	SELECT * WHERE \{ ?person ~nyt:latest\_use ~?mentionInNYT .\\[-1mm]
	\hspace*{1mm} ?person ~owl:sameAs ~?chancellor . ~ ?chancellor ~dct:subject \\[-1mm]
\hspace*{17mm} {\tiny \textless}http://dbpedia.org/resource/Category:Chancellors\_of\_Germany{\tiny \textgreater} \}
\\[-2mm]
\end{scriptsize}

\noindent
We can use the URI at the end of this query as a starting point for a tra\-vers\-al-based execution of the query (under $\cMatch$-bag-se\-man\-tics; cf.~Section~\ref{SecInText:cMatch}).
	For this execution we can use a randomized traversal strategy; that is, we prioritize the retrieval of Linked Data by using randomly chosen lookup priorities for all URIs that are discovered and need to be looked up during the execution process.
By repeating this query execution five times, for each of these executions, we measure an overall execution time of
	8.9
~min (because all five executions eventually retrieve the same set of documents, which always requires almost the same amount of time). However,
	\removable{due to the random prioritization, the documents always arrive in a completely different order. As a consequence, the time after which all data required for computing a particular element of the result set has been retrieved differs for any of the five executions.} These differences have a significant impact on the response times:
In the best of the five cases, a first element of the result set
	can be returned
%
	9~sec, that is, 1.7\%
of the overall query execution time; on average however the five executions require
	3.1~min~(34.8\%)
to return a first result element, and the standard deviation of this average is as high as
	1.3~min~(14.6\%).
Perhaps even more remarkable, the query result is already complete after
	5.2
~min (58.3\%) in the best execution, whereas the worst execution requires
	8.7
~(98.0\%); the corresponding average is
	7.1
~min (80.0\%) and the standard deviation is
	1.2
~min (13.6\%).
\end{example}


The example shows that there exists a huge potential for optimizing the response times of tra\-vers\-al-based queries over the Web of Linked Data (i.e., returning result elements as early as possible).
	In this paper we
propose
	an
approach to realize this potential.

	The basic idea of this approach is to prioritize the lookup of URIs such that
		re\-sult-rel\-e\-vant documents, whose data can be used to compute at least one of the elements of the query result, are retrieved as early as possible. Then, as soon as these documents arrive, a pipelined result construction process can compute and output result elements.
The primary challenge in this context is that the URIs to be looked up are discovered only recursively, and we cannot assume any a priori information about what URIs will be discovered and
	which of the discovered URIs allow us to retrieve documents that are re\-sult-rel\-e\-vant.
We cannot even assume complete, a priori information about the \emph{current} link structure of the queried Web~(i.e., the particular Web graph as it exists at the time of query execution).
Given these issues, we require approaches to prioritize URI lookups that are applicable in the context of tra\-vers\-al-based query execution
	and, for these approaches, we have to identify their suitability for response time optimization.
	To this end, we
make the following main contributions in this paper:

\begin{enumerate}
	\item
		We present
			an \emph{approach to implement} the general idea of \emph{tra\-vers\-al-based query execution} in a way
		that is sufficiently flexible to test any of our ideas for prioritizing URI lookups and their impact on response times (Section~\ref{sec:Implementation}).

	\item
		We show experimentally that \emph{optimization techniques that focus solely on the process of constructing the query result
			do not reduce the
			response times or the overall execution time}
		of tra\-vers\-al-based query executions (Section~\ref{sec:Evaluation1}). Hence, this contribution
			provides evidence
		that the response times of tra\-vers\-al-based queries can be optimized only based on techniques that take the data retrieval process into account.

	\item
		We introduce a diverse set of \emph{%
			\PaperVersion{14 }%
			\TechReportVersion{16 }%
		different approaches to prioritize URI lookups} for
			the data retrieval process of a tra\-vers\-al-based query execution.
		None of these approaches assumes any a priori information about the queried Web (Section~\ref{sec:Approaches}).

%

	\item
		We conduct an experimental analysis to \emph{study the effects} that each of these prioritization approaches can have \emph{on the response times of tra\-vers\-al-based query executions}%
			, and we show that some of the approaches can achieve significant improvements over the baseline approach that looks up URIs on a first-come, first-served basis (Section~\ref{sec:Evaluation}).
\end{enumerate}

	In the remainder of this paper we refer to queries over the Web of Linked Data as \emph{Linked Data queries}. Before discussing our tra\-vers\-al-based approach to execute such queries,
we
	briefly review
the state of the art in
	querying the Web of Linked~Data.
%

%
%

\section{Linked Data Query Processing}
\label{sec:RelWork}

	This section begins with a brief, informal overview on the
theoretical foundations of Linked Data
	queries
(for a more rigorous, formal treatment refer to our earlier work~\cite{Hartig12:TheoryPaper}). Thereafter, we~discuss
	\removable{existing work on}
approaches to execute
	such queries
and elaborate more on the
	focus of the work that we present
in this paper.

\subsection{Foundations}

%
\PaperVersion{%
The prevalent query language used in existing work on Linked Data query processing is the basic fragment of SPARQL that consists of sets of triple patterns. 
Results of this form of conjunctive queries are sets~(or multisets) of \emph{mappings} that associate variables with RDF terms. If the domain of such a mapping is the set of all variables in a given set of triple patterns and replacing their variables according to the mapping produces a set of RDF triples that is a subset of the queried data, then the mapping is called a \emph{solution} for these triple patterns over the data. The complete \emph{query result} of a set of triple patterns over some RDF data consists of all solutions that can be found in the data. A mapping that is a solution only for a proper subset of the triple patterns in a query is called an \emph{intermediate solution}, and any RDF triple that can be obtained by replacing the variables in a single triple pattern according to such an intermediate solution is called a \emph{matching triple} for that triple~pattern%
	\hidden{~(i.e., the matching triple agrees with the pattern on all elements but the~variables)}%
.
} 

\TechReportVersion{%
The prevalent query language used in existing work on Linked Data query processing is a basic fragment of SPARQL, which is the standard language for querying collections of RDF data~\cite{Harris13:SPARQL1_1Language}. This basic fragment consists of
sets of triple patterns. Informally, such a \emph{triple pattern} is an RDF triple whose subject, predicate, or object may be a query variable. Any variable that appears in multiple triple patterns of a query can be conceived of as a join variable. For instance, the query in Example~\ref{ex:Motivation} consists of three triple patterns with two join variables
	({\small ?person}, {\small ?chancellor}).
%
%
%
%
Results of this form of conjunctive queries are sets~(or multisets) of \emph{mappings} that associate variables with RDF terms~(i.e., URIs, blank nodes, or literals). If the domain of such a mapping is the set of all variables in a given set of triple patterns and replacing their variables according to the mapping produces a set of RDF triples that is a subset of the queried data, then the mapping is called a \emph{solution} for these triple patterns over the data. The complete \emph{query result} of a set of triple patterns over some RDF data consists of all solutions that can be found in the data. A mapping that is a solution only for a proper subset of the triple patterns in a query is called an \emph{intermediate solution}, and any RDF triple that can be obtained by replacing the variables in a single triple pattern according to such an intermediate solution is called a \emph{matching triple} for that triple pattern~(i.e., the matching triple agrees with the pattern on all elements but the~variables).
} 

	Multiple proposals exist for adapting the standard query semantics of SPARQL (which focuses on centralized collections of RDF data) to provide for well-defined Linked Data queries~\cite{Bouquet09:QueryingWebOfData, Hartig12:TheoryPaper, Harth12:CompletenessClassesForLDQueries, Umbrich12:LinkedDataQueriesWithReasoning:Article}.
%
	\PaperVersion{A possible approach is a \emph{full-Web query semantics}. Informally, the scope of evaluating a SPARQL query under this semantics }%
	\TechReportVersion{The most common approaches are a full-Web query semantics and several reach\-abil\-i\-ty-based query semantics. Informally, the scope of evaluating a SPARQL query under \emph{full-Web semantics} }%
is the (virtual) union of \emph{all} Linked Data in all documents on the Web. Unsurprisingly,
	\TechReportVersion{the computational feasibility of such queries is very limited; that is, }%
there cannot exist any approach to execute these queries that guarantees both termination and completeness~\cite{Hartig12:TheoryPaper}.
As a computationally feasible alternative~\cite{Hartig12:TheoryPaper}, \emph{reach\-abil\-i\-ty-based query semantics} for SPARQL restrict the scope of any query to
	%
	%
	%
	a que\-ry-spe\-cif\-ic \emph{reachable subweb}%
		\PaperVersion{. To this end, }%
		\TechReportVersion{ consisting only of documents that can be reached by following particular data links. To define this subweb, }%
	the specification of any query in this context includes a set of seed URIs (in addition to the query pattern).
	Then, a document in the queried Web is defined to be \emph{reachable} (and, thus, part of the reachable subweb) if it can be retrieved by looking up either a seed URI---in which case we call it a \emph{seed document}---or a URI that \enumA~occurs in an RDF triple of some other reachable document and \enumB~meets a particular \emph{reachability condition} specified by the given reach\-abil\-i\-ty-based query semantics. For instance, such a condition may require that the triple in which the URI is found is a matching triple for any of the triple patterns in the given query. Our earlier work formalizes this condition in a reach\-abil\-i\-ty-based query semantics that we call \emph{$\cMatch$-se\-man\-tics}~\cite{Hartig12:TheoryPaper}%
		. We shall use this semantics in this paper.
\label{SecInText:cMatch}

\subsection{Query Execution}
Approaches to execute Linked Data queries can be classified
	\PaperVersion{into tra\-vers\-al-based, in\-dex-based, }%
	\TechReportVersion{broadly into tra\-vers\-al-based approaches, in\-dex-based approaches, }%
and hybrid~\cite{Hartig13:Survey, Ladwig10:LinkedDataQueryProcessingStrategies}. All these approaches compute a query result based on Linked Data that they retrieve by looking up URIs during the query execution
	process.
Their strategy to select these URIs is where the approaches~differ.

\emph{Tra\-vers\-al-based approaches} use a set of seed URIs to perform a recursive URI lookup process during which they incrementally discover further URIs that can be selected for lookup.
	Such a selection of further URIs may be based on the aforementioned reachability condition of some reach\-abil\-i\-ty-based query semantics%
		. Hence, tra\-vers\-al-based query execution approaches can support naturally any reach\-abil\-i\-ty-based query semantics.
While existing work on tra\-vers\-al-based query execution
	studies techniques to implement this approach~\cite{%
		Hartig11:HeuristicForQueryPlanSelection, %
		Hartig13:SQUINDemo, %
	Hartig09:QueryingTheWebOfLD, Ladwig11:SIHJoin, Miranker12:Diamond}, in this paper we focus on query optimization, which is a largely unexplored topic in this context.

\emph{In\-dex-based approaches} assume a pre-pop\-u\-lat\-ed index whose entries are URIs, each of which can be looked up to retrieve a Linked Data document. Then, for any possible query, such an approach uses its index to select a set of URIs whose lookup will result in retrieving que\-ry-rel\-e\-vant
	data. By relying on their index, in\-dex-based approaches fail to exploit que\-ry-rel\-e\-vant data added to indexed documents after building the index, and they are unaware of new documents. Existing work on such approaches focuses on the pros and cons of different ways to construct the corresponding
index~\cite{Umbrich11:DataSummariesForLDQueryProcessingArticle}, on techniques to leverage such an index~\cite{Harth10:DataSummariesForLDQueryProcessing}, and on ranking functions that
	prioritize the lookup of
the selected URIs in order to optimize response times~\cite{Harth10:DataSummariesForLDQueryProcessing, Ladwig10:LinkedDataQueryProcessingStrategies}. The latter aims to achieve the same objective as our work in this paper. However, the
	functions to rank URIs as
proposed for
in\-dex-based
approaches rely on statistical metadata that has been added to
the index. For our work on optimizing tra\-vers\-al-based query executions we do not assume an a priori availability of any metadata whatsoever.

The only \emph{hybrid approach} that has been proposed in the literature so far
	performs a tra\-vers\-al-based execution using a prioritized list of URIs to look up~\cite{Ladwig10:LinkedDataQueryProcessingStrategies}. To obtain the initial (seed) version of this list the approach exploits a pre-populated index. Additional URIs discovered during the execution are then integrated into the list, for which they need to be assigned a lookup priority.
To this end,
URIs that are in the index (but have not been selected initially) are prioritized based on a ranking function that uses information from the
	index.
%
	For any URI for which no index entry exists,
the approach simply uses as priority the number of retrieved Linked Data documents that mention the URI in some of their RDF triples (i.e., the number of known incoming links). One of the prioritization approaches that we analyze in this paper resembles this particular strategy for prioritizing non-indexed URIs~(cf.~Section~\ref{ssec:Approaches-PurelyGraphBased}).

We emphasize that all existing work on Linked Data queries assumes
	that the Web of Linked Data does not change
during the execution of a query. We
	make the same assumption
in this paper.
\subsection{Focus of Our Work}
\label{ssec:Contributions}

As discussed in the previous section, the prioritization of URI lookups is an idea that has been shown to be suitable to improve the response time of queries over the Web of Linked Data. However, the only systematic analyses of approaches that implement this idea focus on
	in\-dex-based query executions~\cite{Harth10:DataSummariesForLDQueryProcessing, Ladwig10:LinkedDataQueryProcessingStrategies},
and the approaches proposed in this context cannot be used for a tra\-vers\-al-based execution because they rely on statistical metadata
that
	may be recorded when building an index but that
is not a priori available to a~(pure) tra\-vers\-al-based query execution system~%
	(which also rules out these approaches for non-indexed URIs in a hybrid system).

Therefore, we study the problem of how to prioritize URI lookups in order to optimize response times of tra\-vers\-al-based query executions. In particular, we are interested in the impact that different prioritization approaches have on response times.

The queries that we focus on are reach\-abil\-i\-ty-based, conjunctive Linked Data queries (represented by a set of triple patterns) whose set of seed URIs are the URIs in all triple patterns of the query. For these queries we assume the aforementioned $\cMatch$-se\-man\-tics%
, which is one of the most prominent reach\-abil\-i\-ty-based query semantics supported by the tra\-vers\-al-based approaches studied in the literature~%
	\cite{Hartig11:HeuristicForQueryPlanSelection, Hartig09:QueryingTheWebOfLD, Ladwig11:SIHJoin, Miranker12:Diamond}.
While, in theory, there exist an infinite number of other reach\-abil\-i\-ty-based query semantics and our experiments can be repeated for any of them, we conjecture that the results will be similar to ours because none of the approaches studied in this paper makes use of any specific characteristics of $\cMatch$-se\-man\-tics.
Furthermore, for our study we use the bag version of
	$\cMatch$-se\-man\-tics, which allows us to focus on a notion of the response time optimization problem that is isolated from the additional challenge of avoiding the discovery of duplicates 
(which is an additional aspect of response time optimization under set semantics and worth studying as an extension of our work).

As a final caveat before going into the details, we emphasize~that our study ignores factors that may impact the response times of Linked Data queries but that cannot be controlled by a system that executes such queries~(e.g., the possibility of varying latencies when accessing different Web servers). Moreover, our study focuses only on approaches that do \emph{not} assume any a priori information about the queried Web of Linked Data; that is, the topology of the Web or statistics about the data therein is unknown at the beginning of any query execution. Hence, this focus also excludes approaches that aim to leverage such information collected during earlier query executions~(of course, studying such approaches is an interesting direction for future work).
Similarly, we ignore the possibility to cache documents for subsequent query executions. While caching
	can reduce the time to execute subsequent queries\hidden{ by avoiding some URI lookups}~\cite{Hartig11:EvaluationOfCaching},
this reduction comes at the cost of potentially outdated results. However, studying approaches to balance the performance vs.~freshness trade-off in this context is another interesting direction for future~work.


\section{Implementation Approach} \label{sec:Implementation}

\begin{figure}[t]
	\centering
	\includegraphics[width=0.82\columnwidth]{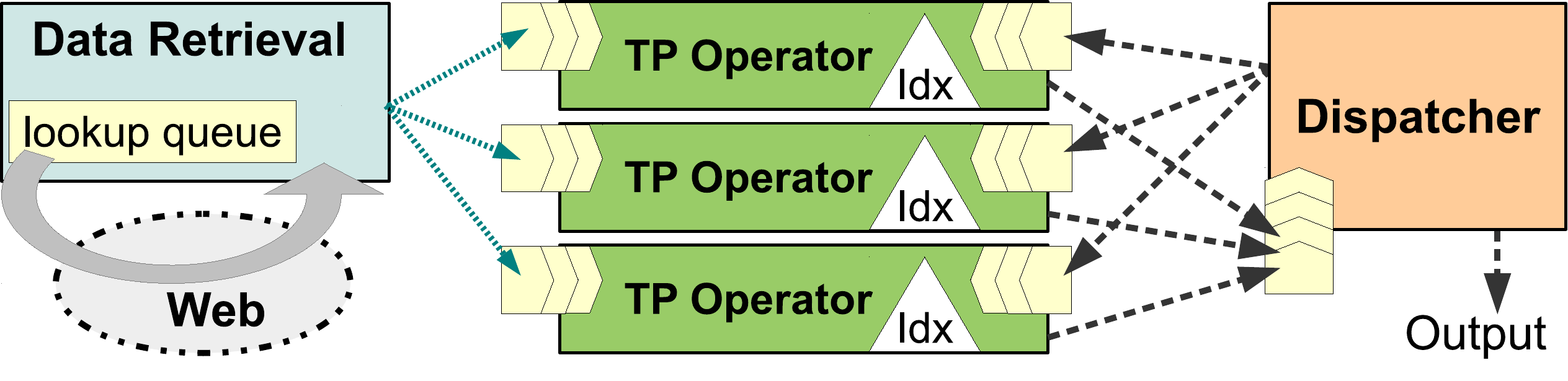}
\PaperVersion{\vspace{-2mm}} 
	\caption{Network of operators as used by our approach to implement the tra\-vers\-al-based query execution paradigm. } 
	\label{fig:operators}
\PaperVersion{\vspace{-3mm}} 
\end{figure}

This section describes our implementation approach to tra\-vers\-al-based query execution.
	In contrast to other approaches that have been proposed in the literature~\cite{Hartig09:QueryingTheWebOfLD, Ladwig11:SIHJoin, Miranker12:Diamond}, our approach
		\TechReportVersion{\removable{adopts ideas of adaptive query processing and} does not rely on any form of static query execution plans\removable{. As a result, our approach} }%
		\PaperVersion{does not rely on any form of static query execution plans and }%
	is flexible enough to study a variety of optimization techniques%
		\TechReportVersion{, some of which cannot be implemented based on static query execution~plans}%
	. \par For any given conjunctive Linked Data query, our approach
uses three types of operators: a data retrieval operator (\drop), a dispatcher, and a set of triple pattern operators~%
	(\tpops)---one for each triple pattern in the query.
Each of these operators is run by a separate thread, and the operators are connected by
	\PaperVersion{message }%
	\TechReportVersion{\Todo{message}{In the description of the experimental setup, clarify that our system, as used for the experiments, uses unbounded FIFO queues, and argue that unboundedness is not an issue / potential bottleneck because of the dominance of data retrieval.} }%
queues as illustrated in Figure~\ref{fig:operators}~(for a query with three triple patterns).
%
%
The operators execute the query using the following process. \par The \drop\ retrieves Linked Data by looking up URIs and sends any retrieved RDF triple that matches a triple pattern of the query to the corresponding \tpop. Given such a matching triple, the \tpop\ generates an initial version of an intermediate solution that
	\removableAltOff{contains variable bindings reflecting}{reflects}
the matching triple; the \tpop\ adds this intermediate solution to an op\-er\-a\-tor-in\-ter\-nal index and passes it to the dispatcher. The dispatcher sends any such initial intermediate solution successively to all other \tpops, each of which probes its own index of initial intermediate solutions to find join
	\removableAltOff{candidates whose variable bindings can be used to complete the incoming intermediate solution before sending it}{candidates. Joined intermediate solutions are then send}
back to the dispatcher. Whenever the dispatcher receives an intermediate solution that has been contributed to by all \tpops, the dispatcher outputs it as a solution of the query result.

\TechReportVersion{
The whole query execution process continues until the \drop\ signals the completion of the data retrieval process~(or a user terminates the execution earlier). At this point, the query execution engine starts
	monitoring the threads of the other operators.
Once all of them are idle and all their message queues are empty, the engine stops the execution and deallocates all
	resources used for it.
}

The careful reader may notice
	that our dispatcher operator presents a form of an Eddy operator~\cite{Avnur00:Eddies}.
%
	We emphasize, however, that
in contrast to existing work on Eddies---in which the source of incoming data
does not receive much attention---%
	the optimization techniques proposed in this paper focus on the \drop.
\TechReportVersion{%
We show that Eddies-style optimization opportunities for the process of constructing query results have little effect on the response times of tra\-vers\-al-based query executions. Nonetheless, our approach leverages certain properties of the Eddies-based result construction process. In particular, some of the
	URI prioritization techniques
that we shall study require
a central component (the dispatcher) that sees all intermediate solutions in all their states. Moreover, we benefit from the
	fully pipelined processing which enables systems 
to compute and report any solution of a query result as soon as all
	data that contributes to this solution has
been retrieved.
}
\TechReportVersion{In the following, we describe our operators in more detail.}
\PaperVersion{
\par
In the following, we describe the \drop\ and the dispatcher in more detail. An extended version of this paper~\cite{ExtendedVersion} provides a detailed discussion of \tpops\ and how our approach guarantees
	correct
results under any reach\-abil\-i\-ty-based query semantics.
}

\subsection{Data Retrieval Operator} \label{ssec:Implementation:DROp}

The \drop\ is equipped with a
	\emph{lookup queue}
that is initialized with the seed URIs of the given query. Starting from these seeds, the operator uses multiple \emph{URI lookup threads} to perform a tra\-vers\-al-based data retrieval process. Whenever such a thread is
	free,
it obtains the
	next
URI from the queue, looks up this URI~on the Web, and scans the RDF triples that are contained in the document retrieved by the lookup.
This scan has the following two~goals.

First, the triples may contain new URIs that can be scheduled for lookup. However, the lookup threads do not necessarily have to add all new URIs to the lookup queue. Instead, the \drop\ can support an arbitrary reach\-abil\-i\-ty-based query semantics by scheduling only those URIs for lookup that satisfy the
	reachability criterion specified by the semantics.
By doing so, the lookup threads
	incrementally discover (and retrieve) the specific reachable subweb that the given query semantics defines as the scope of the query.~%
Hence, all
	RDF triples scanned 
by the lookup threads---and only these---have to be considered to compute the sound and complete query result.
Consequently, the second goal of scanning
	these triples is to identify every triple that matches
a triple pattern in the given query; then, the lookup threads place any such matching triple into the input queue of the corresponding \tpop%
	\TechReportVersion{, which starts processing this triple as soon as it arrives}%
.
If a triple matches multiple triple patterns, it is forwarded to all relevant \tpops.

After retrieving and processing all documents from the que\-ry-spe\-cif\-ic reachable subweb, the \drop\ 
signals the completion of
	the data retrieval process
to the system.
\PaperVersion{At this point (or when a user terminates the execution earlier), the query execution engine starts monitoring the threads of the other operators. Once all of them are idle and all their message queues are empty, the engine stops the execution and deallocates all resources used for it.}

Example~\ref{ex:Motivation} shows that a fundamental aspect of
	such a tra\-vers\-al-based
data retrieval process is the prioritization of URI lookups. A~trivial approach to prioritize URI lookups is to treat all URIs equal and
perform their lookups on a first-come, first-served basis. This approach can be implemented by maintaining the lookup queue of the \drop\ as a FIFO queue%
	\PaperVersion{ and, thus, it resembles a breadth-first traversal strategy}%
.
	The primary hypothesis of this paper is that we may improve the response times of tra\-vers\-al-based query executions by using a prioritization approach that is more sophisticated than
		\PaperVersion{such a }%
		\TechReportVersion{this }%
	trivial, FIFO-based approach. Hence, we consider this approach as our \emph{\baseline} approach and introduce more sophisticated approaches in Section~\ref{sec:Approaches}; these approaches assume that the lookup queue is maintained as a priority~queue.


\TechReportVersion{
\subsection{Triple Pattern Operators} \label{ssec:Implementation:TPOp}

	Each \tpop\ has
two input
	queues---one is connected to the \drop, the other to the dispatcher.
%
%
	For every matching triple that a \tpop\ receives
from data retrieval, it
	performs four steps: First, the operator
generates a
	set of variable bindings from the matching triple.
		More specifically,
	if $\symTP = \tuple{x_1',x_2',x_3'}$ is the triple pattern of the operator, $t = \tuple{x_1,x_2,x_3}$ is the retrieved RDF triple that matches $\symTP$, $\fctVars{\symTP}$ denotes the set of all variables in $\symTP$, and $\symAllTerms$ denotes the set of all RDF terms (i.e., any URI, literal, or blank node), then the generated set of bindings is a mapping $\mu : \fctVars{\symTP} \rightarrow \symAllTerms$ such that $\mu(x_i') = x_i$ for every variable $x_i' \in \fctVars{\symTP}$.
Second, the operator annotates this mapping with a timestamp that is unique for every mapping generated by any of the \tpops\ used for executing the given query (which requires
	coordinating timestamp generation for each query).
	Timestamping is necessary for later stages of the result construction process, and so is the third step, that is, adding the
timestamped mapping into an op\-er\-a\-tor-in\-ter\-nal index.
	\removable{We shall see that} this
index enables the operator to efficiently
	find join candidates that have a given variable binding.
Hence, index keys are pairs of a variable and an RDF term.
In the fourth step, the timestamped mapping is augmented with a vector of \emph{Covered} bits and placed into the input queue of the dispatcher. The bit vector contains a bit for each \tpop\ participating in the query execution; the bit indicates whether the operator
	has already processed the mapping. Hence, in any initial intermediate solution only the bit of the generating operator is set.

In addition to matching triples, a \tpop\ receives intermediate solutions generated initially by other \tpops\ (and sent through the dispatcher). For any such incoming intermediate solution, the operator
	identifies possible join candidates by accessing its internal index. 
Any initial intermediate solution in this index qualifies as a join candidate if it satifies two conditions: First, its timestamp is older than the timestamp of the incoming intermediate solution. Second, both intermediate solutions agree on the binding for any variable bound by both; that is, if $\mu$ and $\mu'$ are the mappings of both intermediate solutions, respectively, then $\mu(x) = \mu'(x)$ has to hold for all variables $x \in \fctDom{\mu} \cap \fctDom{\mu'}$.
	After retrieving the join candidates from its index, the operator creates a copy of each of them. For each of these copies, the operator merges into that copy the variable bindings from the incoming intermediate solution, annotates the copy with the timestamp and the vector of Covered bits of the incoming intermediate solution, sets its own Covered bit, and sends the copy as a new intermediate solution back to the dispatcher.
If there does not exist a join candidate in the index, the operator drops the incoming intermediate solution (i.e., such an intermediate solution is not sent back to the~dispatcher).

We emphasize that the aforementioned time\-stamp-related condition for join candidates is necessary to avoid the generation of spurious duplicates and, thus, to guarantee soundness of computed query results. On the other hand,
	completeness is also guaranteed
because if an initial intermediate solution in the index satisfies the second, bind\-ings-related condition but not the timestamp
	condition,
when the dispatcher sends this intermediate solution to the \tpop\ that generated the initial version of the incoming intermediate solution, then that operator will perform the join.

Another interesting property of \tpops\ is that they can be adjusted easily to support set semantics (using them as described before would result in bag semantics). The only change that is needed for set semantics is to drop any matching triple passed over from data retrieval for which there already exists an initial intermediate solution in the op\-er\-a\-tor-in\-ter\-nal index. However, in this paper we focus on bag semantics.

We note that our \tpops\ are similar to the SteMs used in the original Eddies-based systems~\cite{Madden02:ContinuousQueriesWithAnEddy, Raman03:SteMs}. However,
	while these systems assume relational queries, which have a closure property (i.e., objects that are input to queries are of the same type as the results), we focus on SPARQL-based queries which do not have this property. As a consequence,
our \tpops\ are designed to deal with two different types of inputs from two separate input queues (as opposed to a single queue with relational tuples). To process both inputs in parallel, the \tpop\ as implemented in our system uses two separate threads; one
	is responsible for the queue of matching triples coming from data retrieval, the other processes all intermediate solutions from the dispatcher.

While \tpops\ are the only operators in our framework that contribute to the expressiveness of supported queries, it is possible to develop other types of operators (which might involve adding further state information to intermediate solutions) to support more expressive Linked Data queries. However, such an extension is out of scope of this paper.
}

\subsection{Dispatcher} \label{ssec:Implementation:EddyOp}

	\PaperVersion{For each intermediate solution obtained from a \tpop, }%
	\TechReportVersion{The dispatcher presents our notion of an Eddy-style operator. It has an input queue to receive intermediate solutions from the \tpops, and an output queue to send solutions of the query result to the user. For each intermediate solution obtained through its input queue, }%
the dispatcher checks
\PaperVersion{%
	whether all \tpops\ have contributed to it. To this end, the \tpops\ annotate each intermediate solution with a vector of \emph{Covered} bits that indicate which of the \tpops\ have already processed the intermediate solution~\cite{ExtendedVersion}. If all of these bits are set, the \drop\ can send the intermediate solution
}%
\TechReportVersion{
	the Covered bits. If all of these bits are set, the intermediate solution covers all triple patterns of the query and is sent
}%
to the output. Otherwise the dispatcher routes the intermediate solution to a \tpop\ whose Covered bit is not yet set; if there are multiple candidates, the dispatcher uses a routing policy to
	select
the target operator.
Hence, through its routing policy the dispatcher chooses on the fly the specific order in which the
	\tpops\ 
process intermediate solutions, and this order---which resembles the traditional notion of a query execution plan---may change for each intermediate solution.




Due to this flexibility, Eddy-based query processing has been shown to outperform
static query processing approaches \cite{Avnur00:Eddies, Madden02:ContinuousQueriesWithAnEddy, Acosta15:LinkedDataEddies}, which led to
a number of works that apply the idea of Eddies and study routing policies (e.g.,
	\cite{Bizarro05:ContentBasedRouting, Raman03:SteMs, Raman02:PartialResultsWithEddies, Tian03:DistributedEddies}).
%
Certainly, the results of this research can be applied to optimize the
	process of constructing query results
in our
	context
(e.g., reduce the number of intermediate solutions processed). However, in the following, we show that such an optimization does not have any significant impact on the overall performance of tra\-vers\-al-based query executions over the WWW.


\section{Impact of Local Processing} \label{sec:Evaluation1}
In this section we evaluate experimentally the impact that optimization techniques for the result construction process may have on the performance of a tra\-vers\-al-based query execution.
\PaperVersion{In particular, we aim to identify the fraction of query execution times that such optimizations may affect. To this end, in }%
\TechReportVersion{%
	First, we identify the fraction of query execution times that such optimizations may affect. To this end, we describe an initial experiment that studies the data retrieval overhead of tra\-vers\-al-based query execution. Thereafter, we focus on response time improvements that might be achieved by optimizing the result construction process.

	\subsection{Overhead of Data Retrieval}

	In
}%
addition to our regular \drop~(as described in Section~\ref{ssec:Implementation:DROp}), we implemented a \drop\ that caches all retrieved documents. By using this operator, we can conduct experiments in which we execute any test query twice such that the second execution leverages the cache populated during the first execution. Hence, the second execution does not need to retrieve documents and, thus, the difference between the time to complete each of the two executions represents the overhead of data retrieval.%
\TechReportVersion{%
	\footnote{The difference also includes the overhead of caching all retrieved documents. However, by comparing overall execution times of the first executions that populate the cache with execution times of corresponding executions that use the regular, non-caching \drop, we note that the overhead of caching is negligible.}

%

\TechReportVersion{
\begin{figure*}[t]
\centering
	\subfigure[\small Exec.~times (in seconds) for FedBench LD10 over the WWW.]{%
		\includegraphics[height=35mm]{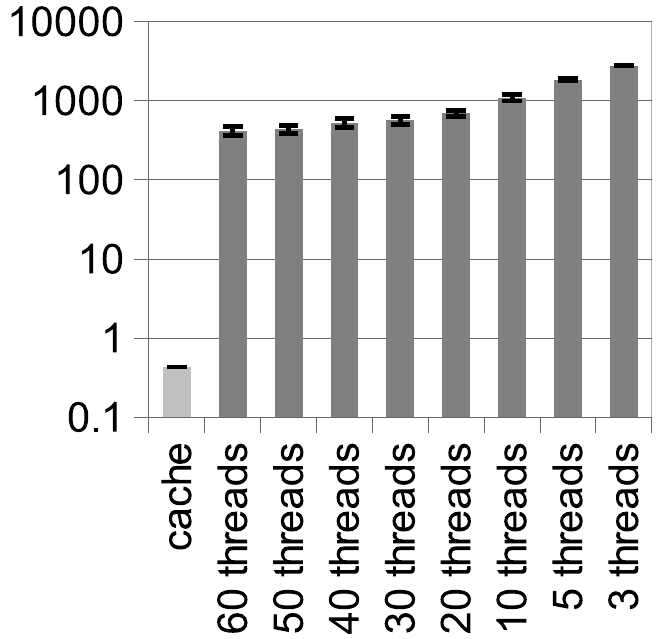}%
		\label{fig:DataRetrievalOverheadFedBenchQ10}
	}
~
	\subfigure[\small Exec.~times (in sec.)~for FedBench over the WWW.]{
		\includegraphics[height=35mm]{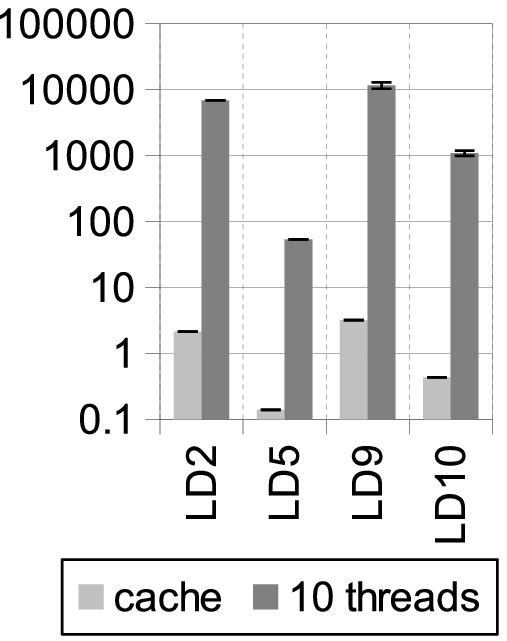}
~
		\label{fig:DataRetrievalOverheadFedBench}
	}
~
	\subfigure[\small \# of intermediate solutions processed by the dispatcher for Q1 over test Web $\symWoD_\mathsf{test}^{62,47}$\!.]{%
		\includegraphics[height=35mm]{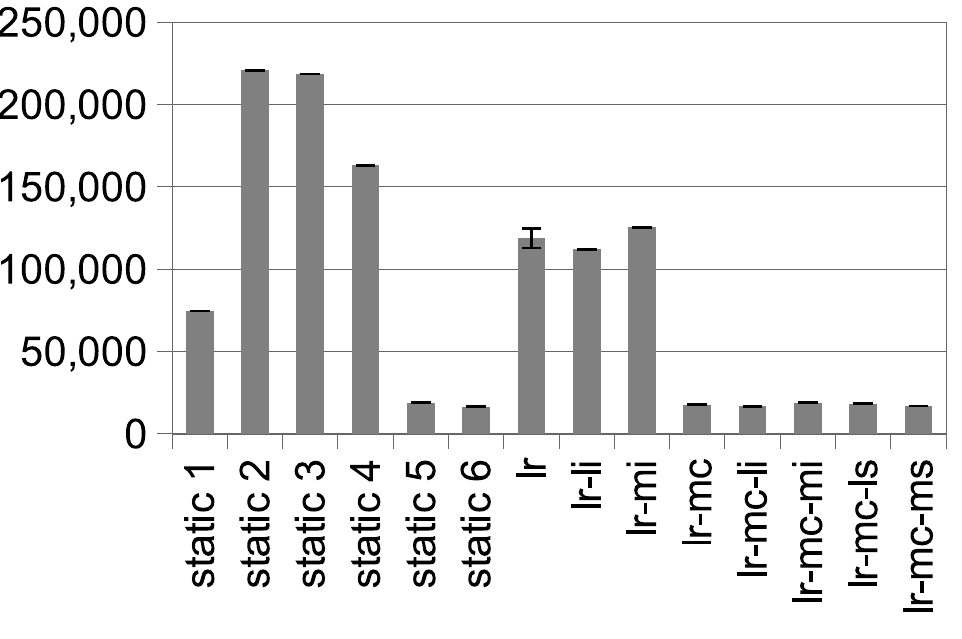}%
		\label{fig:Baseline-IntSolsProcessed-W-62-47-Q1}
	}
~
	\subfigure[\small Relative response times for Q1 over $\symWoD_\mathsf{test}^{62,47}$ (URI prioritization:~baseline).]{%
		\includegraphics[height=35mm]{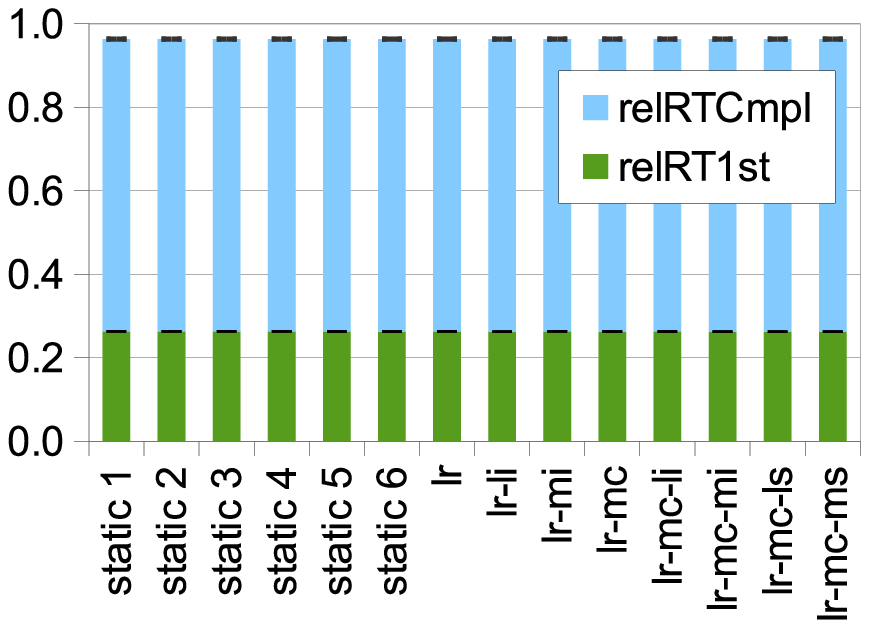}
		\label{fig:Baseline-relRTxxx-W-62-47-Q1}
	}

	\caption{Measurements for optimization techniques that focus on the result construction process.}
\PaperVersion{\vspace{-2mm}} 
\end{figure*}
}

\PaperVersion{
\begin{figure}[t]
\centering
	\subfigure[\small FedBench query LD10.]{%
		\includegraphics[height=35mm]{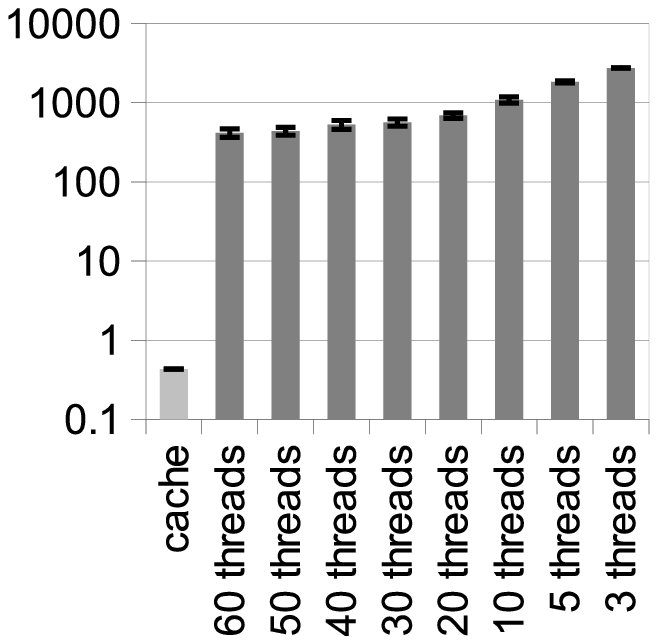}%
		\label{fig:DataRetrievalOverheadFedBenchQ10}
	}
~
	\subfigure[\small Different FedBench queries.]{
		\includegraphics[height=35mm]{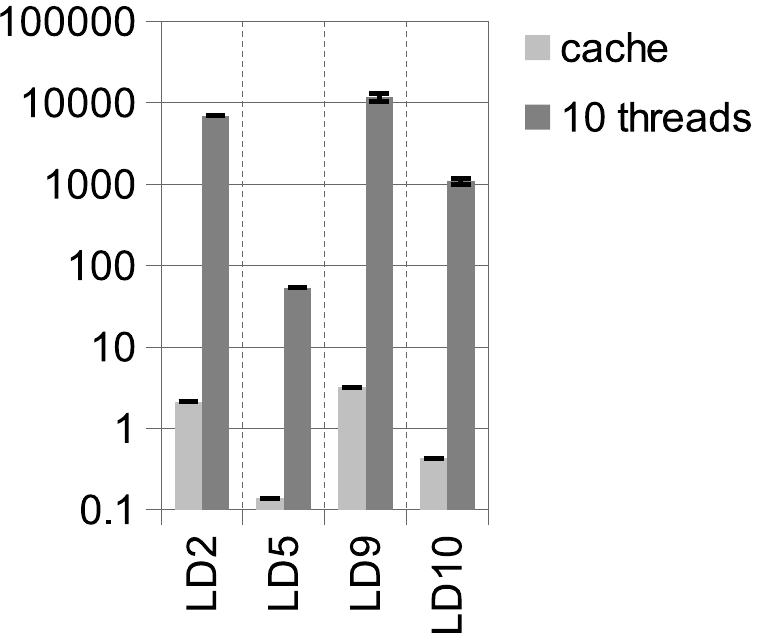}
		\label{fig:DataRetrievalOverheadFedBench}
	}

	\caption{Query execution times (in seconds) over the WWW.}
\end{figure}
}

}
We conducted such an experiment for a number of queries from the FedBench benchmark suite%
	\TechReportVersion{~(cf.~Appendix~\ref{Appendix:FedBenchQueries})}%
, including the query discussed in Example~\ref{ex:Motivation}. For this experiment we configured the dispatcher to route intermediate solutions in a random manner; that is,
	from all \tpops\ whose Covered bit is not set in a given intermediate solution, the dispatcher selects uniformly at random and routes the intermediate solution to the selected operator.
\TechReportVersion{}%
\PaperVersion{

}%
The chart in Figure~\ref{fig:DataRetrievalOverheadFedBenchQ10} illustrates average%
	\TechReportVersion{\footnote{We repeated each query execution five times to obtain the averages. The error bars in the chart represent one standard deviation.}}%
	\PaperVersion{ }%
query execution times measured for the query in Example~\ref{ex:Motivation}%
	\PaperVersion{~(we obtained the averages by repeating each query execution five times; the error bars in the chart represent one standard deviation)}%
.
The leftmost bar in the chart represents the
	cache-based executions~(that do
not need to retrieve documents). The other bars represent executions for which we used our regular
	\drop, which always retrieves all documents from the (que\-ry-spe\-cif\-ic) reachable subweb; we varied the number of parallel lookup threads used by this operator.
These measurements show that data retrieval dominates the overall query execution times by at least three orders of magnitude (note the log scale on the y-axis). We make the same observation for the other queries considered in this experiment; e.g., Figure~\ref{fig:DataRetrievalOverheadFedBench}.
\label{SecInText:DataRetrievalDominance}


A consequence of the
	finding
that data retrieval dominates the overall execution time of tra\-vers\-al-based query executions is that any possible optimization of the local result construction process can only have a negligible effect on these execution times. For the same reason,
\PaperVersion{such optimizations cannot improve response times either~\cite{ExtendedVersion}.}%
\TechReportVersion{%
we assume that such optimizations cannot significantly improve response times either.
	We conducted an experiment to test this hypothesis. In contrast to the aforementioned initial experiment, for this experiment---as well as for any of the other main experiments that we shall describe later in this paper---we use
a simulation environment. Therefore, before we discuss the experiment, we
	specify the experimental setup.
%
All digital artifacts required for our experiments (software, data) are available
	\removableAltOn{on the Web page for the paper}{online}
at~{\small \url{http://squin.org/experiments/ISWC2016/}}.

\subsection{Experimental Setup}

Although the execution of queries over Linked Data on the WWW is the main use case for the concepts discussed in this paper, the WWW is unsuitable as an experimental environment for our
\PaperVersion{analysis. }%
\TechReportVersion{analyses. }%
Reasons for this unsuitability are nondeterministic timeouts, temporarily unresponsive Web servers, unpredictable effects of Web caches, and other, non-re\-peat\-able behavior. Furthermore, certain datasets published as Linked Data on the WWW change frequently; as a consequence, experiments based on queries that discover and use such data quickly become non-re\-pro\-duc\-ible.
Therefore, we set up a simulation environment for our
\PaperVersion{evaluation.}%
\TechReportVersion{experiments.}
	
%

This environment consists of two identical commodity machines, each with an AMD Athlon~64 X2 dual core processor~5600+, 
a
%
	WD
AAKS 500~GB hard disk, and 3.6~GB of main memory.
Both machines use an up-to-date
	\removable{Linux}
Ubuntu~12.04~LTS operating system with Sun Java~1.6.0 and are connected via a fast university network. 
One machine runs an Apache Tomcat Web server~(v.7.0.26) with a Java servlet that can simulate
	different Webs of Linked Data (one at a time); the documents of these Webs
are materialized on the machine's hard disk.
The other machine executes queries over such a simulated Web by using
	our tra\-vers\-al-based
		system, which is an in-memory, Java implementation of the approach in Section~\ref{sec:Implementation}.
%
%
\PaperVersion{%

	For the experiments we configured our system as follows: First, the \drop\ uses a single lookup thread. This configuration allows us to rule out potential effects of parallelized URI lookups as a factor that may influence our measurements. Second, the routing policy that the dispatcher \removable{operator} uses \removable{in the experiments} applies the following selection algorithm: Given an intermediate solution~$\mu$, reduce the set of all candidates (i.e., all \tpops\ for which the Covered bit in $\mu$ is not set) to the operators whose triple pattern contains the least number of variables for which there does not exist a binding in $\mu$. If the resulting set contains multiple candidates, reduce it further by keeping only the \tpop(s) whose internal index has the greatest number of entries at this point. If this still leaves the dispatcher with multiple candidates, it removes all but one by selecting uniformly at random.
	\label{SecInText:SelectedRoutingPolicy}
	We come back to this choice of routing policy in Section~\ref{sssec:Evaluation:Results:ISDriven}.
	In the following, we specify the Webs of Linked Data simulated for our experiments, the corresponding test queries, and the metrics that we use.
}%
\TechReportVersion{%
	For the
		experiments, the \drop\ uses a single lookup thread; this 
	configuration allows us to rule out potential effects of parallelized URI lookups as a factor that may influence our measurements.

	In the following, we specify the Webs of Linked Data simulated for our experiments and the corresponding test queries.
}

\subsubsection{Test Webs}


\begin{table*}[t]
\scriptsize
\centering
\begin{tabular}{|c||c|c|c|c||c|c|c|c|c||c|c|c||c|} \hline
	\cellcolor{gray!50} &
	\multicolumn{4}{c||}{\cellcolor{gray!50}link graph of reachable subweb} &
	\multicolumn{5}{c||}{\cellcolor{gray!50}result-relevant reachable documents} &
	\multicolumn{3}{c||}{\cellcolor{gray!50}res.-irrel.~reachable documents} &
	\cellcolor{gray!50}result
\\ 
	\cellcolor{gray!50}Query &
	\cellcolor{gray!50}\# of &
	\cellcolor{gray!50}\# of &
	\cellcolor{gray!50}str.~connected &
	\cellcolor{gray!50}dia- &
	\cellcolor{gray!50}\# of &
	\cellcolor{gray!50}\% of all &
	\multicolumn{3}{c||}{\cellcolor{gray!50}shortest paths from seed docs} &
	\multicolumn{3}{c||}{\cellcolor{gray!50}shortest paths from seed docs} &
	\cellcolor{gray!50}cardi-
\\ 
	\cellcolor{gray!50} &
	\cellcolor{gray!50}docs &
	\cellcolor{gray!50}edges &
	\cellcolor{gray!50}components &
	\cellcolor{gray!50}meter &
	\cellcolor{gray!50}docs &
	\cellcolor{gray!50}reach.docs &
	\cellcolor{gray!50}mean (st.dev) &
	\cellcolor{gray!50}min &
	\cellcolor{gray!50}max &
	\cellcolor{gray!50}mean (st.dev) &
	\cellcolor{gray!50}min &
	\cellcolor{gray!50}max &
	\cellcolor{gray!50}nality
\\ \hline
Q1 & 3818 & 10007 & 413 & 8 & 572 & 15.0\% & 1.12  ($\pm$0.43) & 1 & 3 & 1.69  ($\pm$0.93) & 1 & 3 & 2481 \\
Q2 & 214 & 627 & 8 & 15 & 22 & 10.3\% & 2.34  ($\pm$1.70) & 1 & 8 & 5.04  ($\pm$1.40) & 2 & 8 & 34 \\
Q3 & 234 & 410 & 57 & 6 & 3 & 1.3\% & 1.41  ($\pm$0.50) & 1 & 2 & 2.74  ($\pm$0.53) & 1 & 3 & 4 \\
Q4 & 1098 & 7805 & 36 & 12 & 43 & 3.9\% & 1.38  ($\pm$0.73) & 1 & 3 & 3.49  ($\pm$0.98) & 1 & 5 & 804 \\
Q5 & 333 & 2340 & 14 & 10 & 36 & 10.8\% & 2.21  ($\pm$0.78) & 1 & 4 & 3.83  ($\pm$0.37) & 3 & 5 & 116 \\
Q6 & 2232 & 6417 & 88 & 45 & 12 & 0.5\% & 2.40  ($\pm$0.78) & 1 & 4 & 4.08  ($\pm$1.34) & 1 & 8 & 28 \\
 \hline
\end{tabular}
\vspace{-2mm} 
\caption{Statistics about
	que\-ry-spe\-cif\-ic reachable subwebs of test Web $\symWoD_\mathsf{test}^{62,47}$\!.}
\label{tab:W-62-47}
\vspace{-2mm} 
\end{table*}

The goal of our experiments is to investigate the impact of different
	\PaperVersion{URI prioritization approaches }%
	\TechReportVersion{optimization techniques }%
on the response times of tra\-vers\-al-based query executions. This impact (as well as the observability thereof) may be highly dependent on how the queried Web of Linked Data is structured and how data is distributed. Therefore, we generated multiple test Webs for our experiments.
To be able to meaningfully compare measurements across our test Webs, we used the same base dataset for generating these Webs.
	\par We selected as base dataset the set of RDF triples that the data generator of the Berlin SPARQL Benchmark (BSBM) suite~\cite{Bizer09:BSBMArticle} produces for a scaling factor of~200.
	This
dataset, hereafter denoted by $\symRDFgraph_\mathsf{base}$,
	consists of 75,150~RDF triples and describes 7,329~entities\TechReportVersion{ in a fictitious e-commerce scenario~(including products, reviews, etc.)}.
Each of these entities is identified by a single, unique URI. Let
	$U_\mathsf{base}$ denote
the set
	consisting of these 7,329~URIs.
Hence, the subject of any base dataset triple $\tuple{s,p,o} \in \symRDFgraph_\mathsf{base}$ is such a URI~(i.e., $s \in U_\mathsf{base}$), and the object $o$ either is a literal or also a URI in $U_\mathsf{base}$.

Every test Web that we generated from this base dataset consists of 7,329 documents, each of which is associated with a different URI in $U_\mathsf{base}$. To distribute the triples of $\symRDFgraph_\mathsf{base}$ over these documents, we partitioned $\symRDFgraph_\mathsf{base}$ into 7,329 potentially overlapping subsets (one for each document).
First, we always placed any base dataset triple whose object is a literal into the subset of the document for the subject of that triple.
Next, for any of the other base dataset triples $\tuple{s,p,o} \in \symRDFgraph_\mathsf{base}$ (whose object $o$ is a URI in $U_\mathsf{base}$), we considered three options:
placing the triple \enumA~into both the documents for $s$ and for $o$---which establishes a bidirectional data link between both documents, \enumB~into the document for $s$ only---which establishes a data link from that document to the document for $o$, or \enumC~into the document for $o$ only---which establishes a data link to the document for $s$.
It is easy to see that choosing among these three options impacts the link structure of the resulting test Web~(note that the choice may differ for each triple).

	We exploited this property to generate test Webs with different link structures. To this end,
we applied a ran\-dom-based approach that, for each generated test Web, uses a particular pair of probabilities $(\phi_1,\phi_2)$ as follows:
For any base dataset triple $\tuple{s,p,o} \in \symRDFgraph_\mathsf{base}$ such that $o \in U_\mathsf{base}$, we chose the first option with probability $\phi_1$; otherwise, $\phi_2$ is the (conditional) probability of choosing the second option over the third.
	To
%
	cover a wide range of possible link structures,
we have used each of the twelve pairs $(\phi_1,\phi_2) \in \lbrace 0,0.33,0.66 \rbrace \times \lbrace 0,0.33,0.66,1 \rbrace$ to generate twelve test Webs $\symWoD_\mathsf{test}^{0,0}, ... \,, \symWoD_\mathsf{test}^{66,100}$\!, and we complemented them with the test Web $\symWoD_\mathsf{test}^{100}$ that we generated using probability $\phi_1=1$%
	\TechReportVersion{~(in which case $\phi_2$ is irrelevant)}%
%
.

While these 13 test Webs cover a wide range of possible link structures, we are also interested in an additional test Web whose link structure is most representative of real Linked Data \removable{on the WWW}. To identify a corresponding pair of probabilities~$(\phi_1,\phi_2)$ we analyzed a comprehensive corpus of real Linked Data that has been crawled for the Billion Triple Challenge 2011~\cite{Harth11:BTC2011Dataset}.
	\PaperVersion{For this corpus, which consists of about 7.9~million \hidden{Linked Data }documents, }%
	\TechReportVersion{This corpus consists of about 7.9~million documents; the overall number of RDF triples distributed over these documents is 2.15~billion (where the overall number of unique RDF triples is 1.97~billion) and the number of URIs mentioned in these triples is 103~million. For this corpus of real Linked Data, }%
we identified a $\phi_1$ of 0.62 and a $\phi_2$ of 0.47. Given this pair of probabilities, we used our base dataset
	\TechReportVersion{$\symRDFgraph_\mathsf{base}$ }%
to generate another test Web, $\symWoD_\mathsf{test}^{62,47}$\!.

We emphasize that
	a systematic creation of test Webs with different link structures as achieved by the given, ran\-dom-based approach 
requires a base dataset that has a high degree of structuredness, which is the case for our
BSBM dataset~\cite{Duan11:ApplesAndOranges}. On the other hand, even if the base dataset is highly structured,
	our ran\-dom-based
approach ensures that
	the documents in each generated test Web~(except $\symWoD_\mathsf{test}^{100}$) contain data with varying degrees of structuredness,
which
	reflects
most of the Linked Data on the~WWW~\cite{Duan11:ApplesAndOranges}.

\subsubsection{Test Queries}

	For our experiments
we use six conjunctive SPARQL queries under $\cMatch$-bag-se\-man\-tics~(cf.~Section~\ref{SecInText:cMatch}). These queries, denoted by Q1 to Q6, are listed in
	\PaperVersion{the extended version of this paper~\cite{ExtendedVersion}. }%
	\TechReportVersion{Appendix~\ref{Appendix:SimulationQueries}. }%

We created these queries so that they satisfy the following three requirements: 
First, each of these queries can be executed over
	any of our test Webs.
%
Second,
	the queries differ w.r.t.~their syntactical structure~(shape, size, etc.).
%
Third, to avoid favoring any particular traversal strategy, the reachable subwebs induced by the queries
	differ from one another along various dimensions%
%
.
%
%
	For instance, Table~\ref{tab:W-62-47}
lists several properties of the six que\-ry-spe\-cif\-ic reachable subwebs of test Web $\symWoD_\mathsf{test}^{62,47}$\!. These properties are the number of reachable documents, the number of edges between these documents in the link graph of the reachable subweb, the number of strongly connected components and the diameter of the link graph, the number of reachable documents that are \emph{re\-sult-rel\-e\-vant} (i.e.,
	their data is required for 
at least one solution of the corresponding query result), the percentage of reachable documents that are re\-sult-rel\-e\-vant, the mean lengths of the shortest paths (in the link graph) from seed documents to these relevant documents, the lengths of the shortest and the longest of these shortest paths, and similar statistics for the reachable documents that are not re\-sult-rel\-e\-vant. Additionally, Table~\ref{tab:W-62-47} lists the cardinality of the corresponding query results.
	We emphasize
that a computation of any of the properties in Table~\ref{tab:W-62-47} requires information that a tra\-vers\-al-based system discovers only during query execution. Hence, such statistics can be computed only after \emph{completing} a tra\-vers\-al-based query execution and, thus, they cannot be used for query optimization.

By comparing the properties in Table~\ref{tab:W-62-47}, it can be
	\TechReportVersion{easily }%
observed that our six test queries induce a
	\TechReportVersion{very }%
diverse set of reachable subwebs of test Web $\symWoD_\mathsf{test}^{62,47}$\!. In
	an earlier
analysis of
	these
queries we make the same observation for
	\TechReportVersion{each of the other 13 }%
	\PaperVersion{the other 13 }%
test Webs%
	~\cite{Hartig14:ReachableSubwebs}%
.
Moreover, if we consider each query in separation and compare its reachable subwebs across the different test Webs, we observe a similarly high diversity~\cite{Hartig14:ReachableSubwebs}.
Hence, these
	\PaperVersion{\removable{six} }%
	\TechReportVersion{six }%
queries in combination with all 14 test Webs represent a broad spectrum of test cases%
	, including some that reflect interlinkage characteristics of a real snapshot of Linked Data on the WWW~(i.e., $\symWoD_\mathsf{test}^{62,47}$) and some that systematically cover
		other possible interlinkage characteristics~($\symWoD_\mathsf{test}^{0,0}$~...~$\symWoD_\mathsf{test}^{100}$).

%

\PaperVersion{%
	\subsubsection{Metrics}
	For each solution that our test system computes during a query execution, it
measures and records the fraction of the overall execution time after which the solution becomes available.
	An example of such numbers for the first and the last reported solutions are the percentages given in Example~\ref{ex:Motivation}.
		For our analysis we will also focus primarily on these two extreme cases---the relative response times for a first solution and the relative response times for the last solution. The former is interesting because it identifies the time after which users can start looking over some output for their query; the latter marks the availability of the complete result (even if the system cannot verify the completeness at this point).
Hence, we define
	the following two main metrics for our analysis.
%
\begin{definition}
	Let \textit{exec} be a query execution; let $t_\mathsf{start}$, $t_\mathsf{end}$, $t_\mathsf{1st}$, and $t_\mathsf{last}$ be the points in time when \textit{exec} starts, ends, returns a first solution, and returns the last solution, respectively.
	The \definedTerm{relative first-so\-lu\-tion response time} (\definedTerm{\textsf{\small relRT1st}}) and the \definedTerm{relative com\-plete-re\-sult response time} (\definedTerm{\textsf{\small relRTCmpl}}) of \textit{exec} are defined as follows:
	\begin{equation*}
		\mathsf{relRT1st} \definedAs \frac{t_\mathsf{1st} - t_\mathsf{start}}{t_\mathsf{end} - t_\mathsf{start}}
		\quad \text{ and } \quad
		\mathsf{relRTCmpl} \definedAs \frac{t_\mathsf{last} - t_\mathsf{start}}{t_\mathsf{end} - t_\mathsf{start}}
		.
	\end{equation*}
\end{definition}

We use
	these fractions
for our analysis because they reveal differences in response times that can be achieved by different
	optimization
approaches \emph{relative to each other}. Measuring absolute times---such as the times that Example~\ref{ex:Motivation} provides in addition to the percentages---would not provide any additional insight for such a comparison. Moreover, absolute times that we may measure in our simulation environment would be quite different from what could be measured for queries over the ``real'' Web of Linked Data~(such as in Example~\ref{ex:Motivation}) because absolute times in our simulation are mostly a function of how fast our simulation server responds to URI lookup requests. 
Relative measurements
	do not have this~problem.

To increase the confidence in our measurements we repeat any query execution five times and report the geometric mean of the measurements obtained by the five executions. The confidence intervals (i.e., error bars) in the charts in this paper represent one standard deviation.
To avoid measuring artifacts of concurrent query executions we execute queries sequentially. Moreover, to also exclude possible interference between subsequent query executions we stop and restart the system between any two executions.

}

\subsection{Effectiveness of Optimizations for the Result Construction Process}

The aim of our first main experiment is to study whether and how response times of a tra\-vers\-al-based query execution are affected by possible optimizations for the local result construction process. To this end, we consider the possibility of using different static execution plans that implement different join orders, as well as the possibility of using a dynamic plan that can change the join order
	for each intermediate solution.
Recall that our implementation of tra\-vers\-al-based query execution enables
	such dynamic plans
by
	employing the dispatcher operator and its routing policy.
%
In the following, we introduce the routing policies considered in this experiment, specify the metrics for our analysis, and discuss the results.

\subsubsection{Routing Policies} \label{ssec:RoutingPolicies}
To specify our routing policies, let $\mu$ be an intermediate solution to be routed and $O$ be the set of \tpops\ whose Covered bit in $\mu$ is not set. Each of the following eight routing policies
requires the dispatcher to first reduce~$O$ to a particular subset $O_x \subseteq O$; then, the dispatcher has to choose uniformly at random some operator from this subset and route $\mu$ to that operator.
	We now specify the routing policies based on these subsets as follows:
\begin{itemize}
%

	\item
		The \emph{least remaining} (\policy{lr}) policy uses a subset $O_\mathsf{lr} \in O$ that includes only the operators whose triple pattern contains the least number of variables for which there does not exist a binding in $\mu$.

	\item
		The \emph{least remaining--least indexed} (\policy{lr-li}) policy is more restrictive than \textsf{lr}; that is, its subset $O_\textsf{lr-li}$ is the subset of $O_\mathsf{lr}$ that is generated by excluding any operator whose internal index of initial intermediate solutions has more entries at this point than the index of some other operator in~$O_\mathsf{lr}$.

	\item
		The \emph{least remaining--most indexed} (\policy{lr-mi}) policy is also more restrictive than \policy{lr}; its subset $O_\textsf{lr-mi}$ is generated by keeping only the operator(s) from $O_\mathsf{lr}$ whose internal index has the greatest number of entries at this point.

	\item
		The \emph{least remaining--most covered} (\policy{lr-mc}) policy is also more restrictive than \policy{lr}; its subset $O_\textsf{lr-mc}$ contains only the operators from $O_\mathsf{lr}$ whose triple pattern has the greatest number of variables for which there exists a binding in $\mu$.

	\item
		The \emph{least remaining--most covered--least indexed} (\policy{lr-mc-li}) policy is more restrictive than \policy{lr-mc}; its set $O_\textsf{lr-mc-li}$ is obtained by
			applying the same reduction to $O_\textsf{lr-mc}$ that \policy{lr-li} applies to $O_\mathsf{lr}$.

	\item
		The \emph{least remaining--most covered--most indexed} (\policy{lr-mc-mi}) policy is also more restrictive than \policy{lr-mc};
		$O_\textsf{lr-mc-mi}$ is obtained by reducing $O_\textsf{lr-mc}$ in the same manner as \policy{lr-mi} reduces~$O_\mathsf{lr}$.

	\item
		The \emph{least remaining--most covered--least selective} (\policy{lr-mc-ls}) policy is also more restrictive than \policy{lr-mc}; its set $O_\textsf{lr-mc-ls}$ contains only the operators from $O_\textsf{lr-mc}$ whose current selectivity is the lowest among all operators in $O_\textsf{lr-mc}$. At any point during the query execution, the \emph{current selectivity} of each \tpop\ is the number of incoming intermediate solutions processed so far by the operator divided by the number of intermediate solutions returned so far by the operator.

	\item
		The \emph{least remaining--most covered--most selective} (\policy{lr-mc-ms}) policy is
			\TechReportVersion{also more restrictive than \policy{lr-mc}; its set $O_\textsf{lr-mc-ms}$ contains only the operators from $O_\textsf{lr-mc}$ }%
			\PaperVersion{%
				also more restrictive than \policy{lr-mc}; its set $O_\textsf{lr-mc-ms}$ contains only the operators from $O_\textsf{lr-mc}$
			}%
		that have the highest current selectivity.
%
\end{itemize}
%
	These routing policies present
a form of \emph{con\-tent-based routing}~\cite{Bizarro05:ContentBasedRouting}.

In addition to dynamic execution plans that emerge from using any of
	the aforementioned
routing policies, we also consider static execution plans in our analysis; these are interesting because any other tra\-vers\-al-based query execution approach proposed in the literature so far is based on static plans~\cite{Hartig09:QueryingTheWebOfLD, Ladwig11:SIHJoin, Miranker12:Diamond}.
To simulate static execution plans in our implementation we use
	\emph{static-plan routing policies}
that enforce a fixed join order for all intermediate solutions. Due to the high number of such join orders, even for queries with
	few
triple patterns, we focus on left-deep join trees only. That is, for a test query with $n$ triple patterns, we use $n!$ such static-plan routing policies, each of which enforces one of the $n!$ left-deep join trees of that query (%
the largest
	of our test queries consists of five triple patterns, i.e., 120 left-deep join trees).


\subsubsection{Experimental Results}

	The charts in Figures \ref{fig:Baseline-IntSolsProcessed-W-62-47-Q1} and \ref{fig:Baseline-relRTxxx-W-62-47-Q1} illustrate measurements obtained by using the different routing policies to execute query Q1 over test Web~$\symWoD_\mathsf{test}^{62,47}$\!.
	Each of the six leftmost bars in these charts presents query executions that use a different static-plan routing policy (there exist six different left-deep join trees for query Q1). Each of the other bars presents query executions that use one of the eight con\-tent-based routing policies, respectively.

Figure~\ref{fig:Baseline-IntSolsProcessed-W-62-47-Q1}
	illustrates the overall number of intermediate solutions processed during these query executions. These numbers show that the amount of work performed by the local result construction process may differ significantly depending on the routing policy used (or on the join order in the case of static execution plans).
However, these differences have no impact on the response times as we can see in Figure~\ref{fig:Baseline-relRTxxx-W-62-47-Q1}. That is, independent of which routing policy it uses, our system always achieves a \textsf{\small relRT1st} of~0.26 and a \textsf{\small relRTCmpl} of 0.99. We observe the same independence for the other five queries, as well as in our other test Webs.
While this independence may seem surprising at first,
	it verifies the aforementioned hypothesis: Optimizing the result construction process of tra\-vers\-al-based query executions has no measurable impact on the response times of these executions. The data retrieval overhead of these executions marginalizes any effect of such optimizations.

}
Therefore, to
	optimize
tra\-vers\-al-based que\-ry executions, we may largely ignore local-processing specific optimization opportunities and, instead, focus on the
	data retrieval process.
A~fundamental issue of this process is the prioritization of URI lookups; Example~\ref{ex:Motivation} demonstrates that this issue needs to be addressed to improve response times of tra\-vers\-al-based query executions.
\TechReportVersion{Consequently, in the remainder of this paper we study approaches to prioritize URIs.}

\section{Approaches to Prioritize URIs} \label{sec:Approaches}

\begin{figure}[tb]
	\centering
	\PaperVersion{\includegraphics[width=0.477\textwidth]{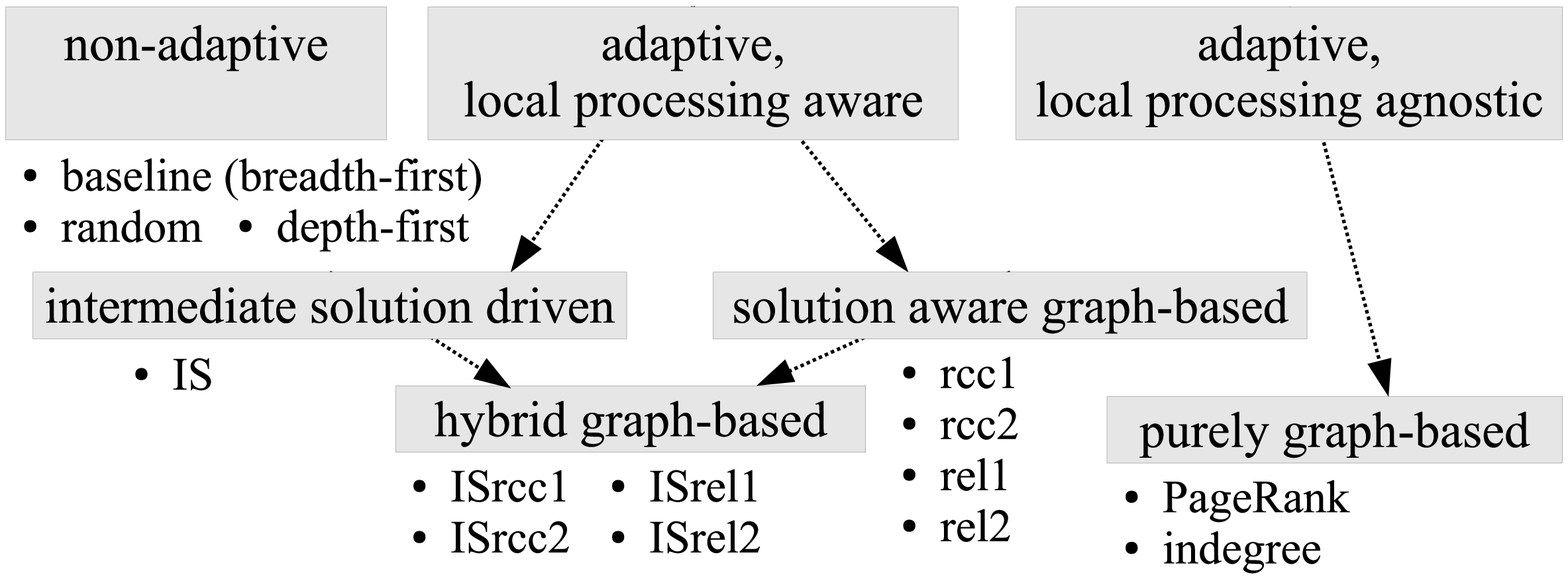}}%
	\TechReportVersion{\includegraphics[width=0.477\textwidth]{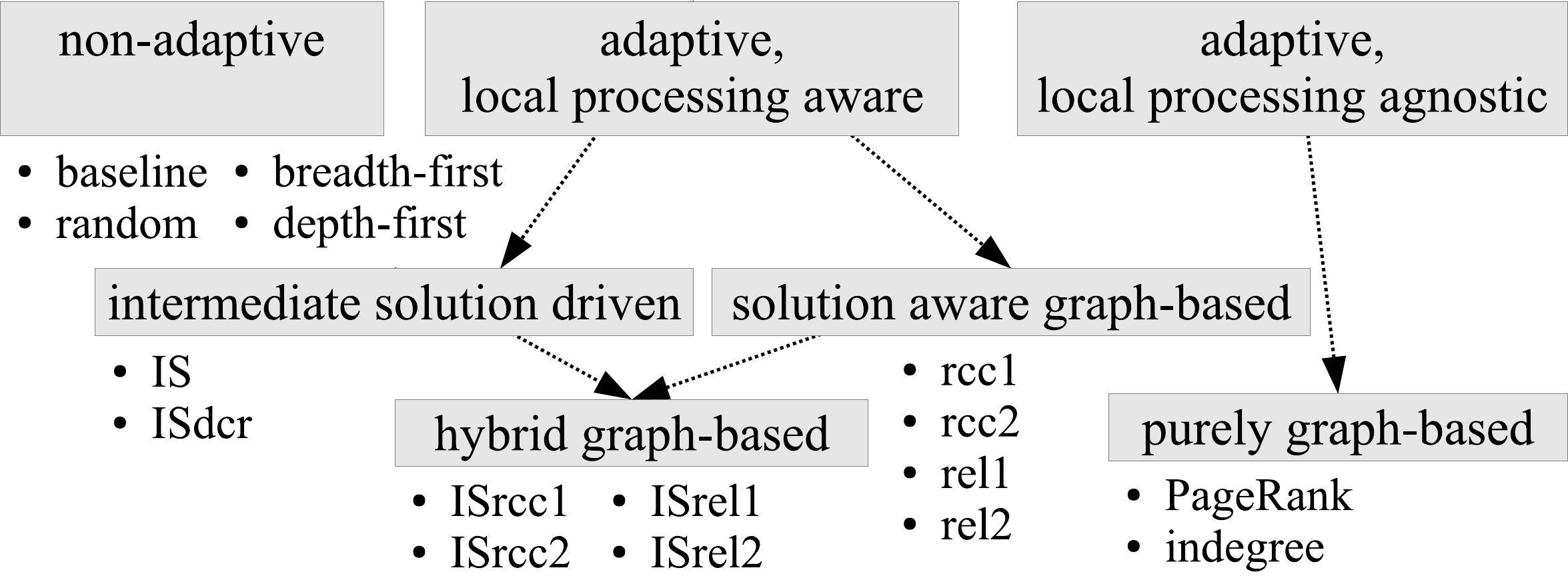}}%
\PaperVersion{\vspace{-2mm}} 
	\caption{Taxonomy of approaches to prioritize URI lookups.}
	\label{fig:taxonomy}
\PaperVersion{\vspace{-2mm}} 
\end{figure}

	A variety of approaches to prioritize URI lookups
are possible.
	In this section, we identify different classes of such approaches~(cf. Figure~\ref{fig:taxonomy}). Thereafter, in Section~\ref{sec:Evaluation}, we
		evaluate their suitability to reduce the response times of tra\-vers\-al-based query executions.

A first class includes approaches that determine a \emph{fixed} priority for each URI when the URI is added to the lookup queue. Since these approaches do not change priorities of queued URIs, we call them \emph{non-adap\-tive approaches}.
\PaperVersion{%
	The \baseline\ approach introduced in Section~\ref{ssec:Implementation:DROp} is a trivial example of a non-adap\-tive approach.
		Recall that this approach resembles a breadth-first traversal strategy. In the extended version of this paper we also discuss a depth-first and a random approach as alternative non-adap\-tive approaches (Example~\ref{ex:Motivation} uses the latter). These turn out to be unsuitable for response time optimization of tra\-vers\-al-based query executions~\cite{ExtendedVersion}.

	\emph{Adaptive ap\-proach\-es}, which may reprioritize queued URIs under specific conditions,
}%
\TechReportVersion{%
	Their opposite, \emph{adaptive ap\-proach\-es},
}%
can be
	divided
further into
	\emph{local processing aware
approaches} that
	take runtime information about the query-local result construction process into account,
and \emph{local processing agnostic approaches} that ignore what happens in the \tpops\ and the dispatcher. A subclass of the latter are approaches that use information about the topology of the queried
	Web, which becomes available gradually during the data retrieval process. Then, these approaches
		repeatedly compute a vertex scoring function such as PageRank over an incrementally constructed model of the Web graph and use the resulting scores as lookup priorities.
We refer to these approaches as \emph{purely graph-based approaches}. A~local processing aware variation of these are \emph{so\-lu\-tion-aware graph-based approaches}, which use a vertex scoring function that leverages information about the solutions of the query result already produced by the result construction process. An alternative subclass of local processing aware approaches are \emph{intermediate solution driven approaches} that utilize information about any intermediate solution that is routed through the dispatcher. Finally, we recognize the possibility to combine the idea of so\-lu\-tion-aware graph-based approaches and intermediate solution driven approaches into \emph{hybrid} local processing aware~approaches.%
\PaperVersion{%
	~In the
		\removableAltOn{remainder of this section}{following,}
	we describe each class in more detail and introduce particular
		\removableAltOn{approaches (which we evaluate experimentally in Section~\ref{sec:Evaluation})}{approaches}.

	All these approaches assume that the lookup queue of the \drop\ is maintained as a priority queue (instead of a FIFO queue as can be used for the \baseline). We denote each priority by a number; the greater the number, the higher the priority. URIs that are queued with the same priority are handled in a first-come, first-served manner (after all higher priority URIs have been looked up).
}%
\TechReportVersion{%

	In the remainder of this section we describe each class in more detail and introduce
		particular approaches
	(which we evaluate experimentally in Section~\ref{sec:Evaluation}).
}

\TechReportVersion{
	\PaperVersion{%
	\subsection{Oracle Approach} \label{ssec:Approaches-Oracle}
}

\TechReportVersion{%

\subsection{Non-Adaptive Approaches} \label{ssec:Approaches-NonAdaptive}

The \baseline\ approach introduced in Section~\ref{ssec:Implementation:DROp} is a trivial example of a non-adap\-tive approach. Recall that this approach can be implemented by maintaining the lookup queue of a \drop\ as a FIFO queue. All other approaches discussed in
	the following
assume a priority queue instead. We denote each priority by a number; the greater the number, the higher the priority. URIs
	that are queued
with the same priority are handled in a first-come, first-served manner (after all higher priority URIs have been looked~up).

A first non-adap\-tive example of such a priority-based approach is \emph{\random}, which, for every URI to be queued, selects uniformly at  random a priority from
	a range of priorities; in this paper we use the interval [1,10] (Example~\ref{ex:Motivation} uses this approach).
%
An alternative~(non-adap\-tive) approach is to start with priority 0 for all seed URIs, and, for any other URI discovered and queued later, the priority is one greater than the priority of the URI whose lookup resulted in the discovery of the URI to be queued. As a result, URIs for (reachable) documents that are farther from the seeds have a higher priority. Hence, by using this approach, a \drop\ traverses
	any queried Web~(more precisely, the query-spe\-cif\-ic reachable portion thereof)
in a depth-first manner. Therefore, we call this approach \emph{\DF}.
Naturally, there also exists a \emph{\BF} approach that resembles a breadth-first traversal.
	This approach works almost the same as the \DF\ approach except that it decreases priorities by one (instead of increasing them).
Hence, any number that denotes a priority in
	the \BF\
approach is either~0 (for the seed URIs) or a negative integer.

} 

To also gain an understanding of what response times a
	tra\-vers\-al-based
system could achieve if it had complete information of the queried Web (which is impossible in practice), we also designed an \emph{\oracle} approach for our experiments. For any given query, this approach assumes an oracle that, for each reachable document, knows \enumA~the URIs leading to the document and \enumB~the
	\PaperVersion{(final) RCC of the document~(i.e., the number of solutions of the complete query result that are based on matching triples from the document). }%
	\TechReportVersion{\emph{result contribution counter} (\emph{RCC}) of the document; that is, the number of solutions of the complete query result that are based on matching triples from the document. }%
\label{SecInText:OracleApproach}
Then, the \oracle\ approach uses as priority of a URI lookup the RCC of the document that will be retrieved by this lookup. As a consequence,
	documents with a greater RCC are retrieved earlier.
Nonetheless,
	for any (non-seed) document with a high RCC, the tra\-vers\-al-based process still has to~discover a URI for the document before the highly prioritized retrieval of the document can be scheduled.
Clearly,
	without a priori information about the queried Web, 
a tra\-vers\-al-based system
%
	cannot determine RCCs before retrieving all reachable documents%
---which is when it is too late to start prioritizing URI lookups. Hence, the \oracle\ approach cannot be used in practice. However, for our experiments we performed a \baseline-based ``dry run'' of our test queries and collected the information necessary to determine the RCCs required
	\PaperVersion{by }%
	\TechReportVersion{to execute the queries using }%
the \oracle~approach.

\TechReportVersion{

Note that none of the
	five aforementioned approaches (\baseline, \random, \DF, \BF, and \oracle)
attempts to reconsider the priority of a URI once the URI is in the lookup queue. In contrast, all approaches that we introduce in the following are adaptive; that is, they reprioritize queued URIs under specific conditions.

} 
}
\subsection{Purely Graph-Based Approaches} \label{ssec:Approaches-PurelyGraphBased}
A first class of adaptive approaches
	to prioritize (and, potentially, reprioritize) URI lookups
is based on
	the idea of applying a vertex scoring method to
a directed graph that represents the topology of the queried Web as discovered during the data retrieval process.

Each vertex in this graph
	corresponds to
either a retrieved document or a queued URI. Each directed edge between two document vertices represents a data link that is established by URIs that occur in some RDF triple in the source document and that turned out to resolve to the target document when looked up. Directed edges from a document vertex to a URI vertex represent data links to documents that are yet to be retrieved.
Obviously, such a graph is an incomplete model of the topology of the queried Web. However, as a side-effect of the data retrieval process, the \drop\ obtains increasingly more information about the topology and, thus, can augment its model continuously. That is, any URI vertex becomes a document vertex after the corresponding URIs has been looked up~%
	(or it is removed if the lookup fails).
If such a lookup results in discovering new URIs for the lookup queue, new URI vertices and connecting edges can be added to the graph. Similarly, new edges can be added if a retrieved document mentions URIs that either
	have been looked up earlier (during the same data retrieval process) or are currently queued for lookup.

Given such a graph, it becomes possible to apply a vertex scoring method and use the score of each URI vertex as the priority of the corresponding URI in the lookup queue. Whenever the graph changes due to the completion of some URI lookup, the scores can be recomputed, and the priorities can be adapted accordingly.
\TechReportVersion{Depending on the vertex scoring method applied, recomputing the scores either needs to be done from scratch or can be done by adjusting previously computed scores incrementally.}

While a multitude of different vertex scoring methods exist, we select PageRank and in\-de\-gree-based scoring as two examples for our study. PageRank is a well-known method that uses an iterative algorithm to determine a notion of importance of vertices, which has been used to rank Web search results~\cite{Page99:PageRank}.
	In\-de\-gree-based~scoring
is a less complex method that simply uses the number of incoming edges as the score of a vertex. Hereafter, we refer to the two resulting URI prioritization approaches as \PageRank\ and \indegree, respectively. The latter approach is equivalent to the only existing proposal to prioritize URI lookups during traversal-based query executions~\cite{Ladwig10:LinkedDataQueryProcessingStrategies}, but its effectiveness has not been studied so far.

\subsection{Intermediate Solution Driven Approaches} \label{ssec:Approaches-ISDriven}

\PaperVersion{%
	We now turn to local processing aware approaches that aim to leverage runtime information about the query-local result construction process.
}%
\TechReportVersion{%
	None of the approaches introduced so far is designed to make use of runtime information about the query-local result construction process; we now turn to local processing aware approaches that aim to leverage such information.
}%
To enable an implementation of these approaches we augment the network of query execution operators as described in Section~\ref{sec:Implementation} with a \emph{feedback channel} from the dispatcher to the \drop. Then, specific information required to prioritize URI lookups can be sent over this channel.
%
%
\PaperVersion{}%
\TechReportVersion{\par}%
\emph{Intermediate solution driven approaches} use the feedback channel to obtain information about intermediate solutions routed by the dispatcher.
\PaperVersion{%
	\par
	We focus on one such approach, called \emph{\is}, that assigns an initial priority of 0 to any new URI added to the lookup queue, and it reprioritizes queued URIs based on the following two assumptions:
}%
\TechReportVersion{%
	In this paper, we focus on two such approaches%
		---\hidden{hereafter, }called \emph{\is} and \emph{\isDecr}---%
	that are based on the following two assumptions:
}%

\begin{enumerate}
\item[A1:]
	The more triple patterns a given intermediate solution covers, the more likely it is that this intermediate solution can be completed into a solution that covers the whole query.
\item[A2:]
	Documents that are most likely to contain (matching) triples for completing a given intermediate solution into a solution are the documents that can be retrieved by looking up the URIs mentioned in the intermediate solution.
\end{enumerate}

Given that the overall optimization objective is to return
solutions as soon as possible, due to
	assumption A2,
it seems reasonable to
	increase the priority of a URI in the lookup queue
if the URI is mentioned in an intermediate solution. Furthermore, due to the
	assumption A1,
such an increase should be greater if the intermediate solution covers more triple patterns.
Consequently,
	\PaperVersion{the \is~approach performs }%
	\TechReportVersion{both the \is~approach and the \isDecr\ approach perform }%
the following two-step procedure for every intermediate solution routed by the dispatcher: First, count the number of Covered bits that are already set in the intermediate solution. Let $cnt$ be this number. Second, iterate over all variables that are bound by the intermediate solution. For each such variable, if the intermediate solution binds the variable to a URI and this URI is queued for lookup with a priority value that is smaller than $cnt$, then increase the priority of this URI to $cnt$.

\TechReportVersion{%
After describing how both approaches, \is\ and \isDecr, reprioritize queued URIs, it remains to specify how these approaches determine the initial priority
	of any new URI that is added to the lookup~queue.
The \is\ approach simply assigns an initial priority of 0 to all new URIs. In contrast, the \isDecr\ approach uses the same strategy as the \BF\
	approach;
that is, all seed URIs start with a priority of 0, and any other URI
	is queued 
with an initial priority that is one less than the (final) priority of the URI whose lookup resulted in the discovery of the URI to be queued. Hence, if the \isDecr\ approach increases the priority of a URI lookup as described above, then the approach also gives all URIs discovered by this lookup a head~start.
} 

\subsection{Solution-Aware Graph-Based Approaches} \label{ssec:Approaches-RCCGraphBased}

An alternative class of local processing aware approaches revisits the idea of graph-based approaches (cf.~Section~\ref{ssec:Approaches-PurelyGraphBased}). Given the possibility to obtain runtime information from the dispatcher,
it becomes feasible to use vertex scoring methods that leverage such information%
	\TechReportVersion{~(as opposed to vertex scoring methods that are purely graph-based such as PageRank)}%
.
In this paper we focus on methods that are based on the number of solutions that
	\TechReportVersion{specific sets of }%
retrieved documents have contributed~to.


To
	\PaperVersion{this end, }%
	\TechReportVersion{enable the application of such methods, }%
intermediate solutions must be augmented with provenance metadata. In particular, each intermediate solution must be annotated with a set of all documents that contributed a matching triple to the construction of that intermediate solution. To this end, before sending a matching triple to a \tpop, the \drop\ augments this triple with metadata that identifies the source document of the triple. This document becomes the provenance of the initial intermediate solution that the corresponding \tpop\ generates from the matching triple. When a \tpop\ joins two intermediate solutions%
	\TechReportVersion{(as described in Section~\ref{ssec:Implementation:TPOp})}%
, the union of their
	provenance annotations
becomes the
provenance of the resulting intermediate~solution.
Then, whenever the dispatcher obtains an intermediate solution that has all Covered bits set (and, thus, is a solution that can be sent to the output), the dispatcher uses the feedback channel to send the provenance annotation of that intermediate solution to the \drop. 
\PaperVersion{}%
\TechReportVersion{\par}%
The \drop\ uses these annotations to maintain a \emph{result contribution counter}~(\emph{RCC}) for every document vertex in the link graph that the operator builds incrementally as described in Section~\ref{ssec:Approaches-PurelyGraphBased}. This counter represents the number of solutions that the document represented by the vertex
has contributed to so far%
	\hidden{, which may increase \removable{monotonically} as the query execution~continues}%
.

Given these counters, we define four vertex scoring functions which can be applied to the link graph.
%
%
Informally, for each vertex $v \in V$ in such a graph $G=(V,E)$, the \emph{rcc-1 score} of $v$, denoted by $\mathrm{rccScore}^1\!(v)$, is the sum of the (current) RCCs of all document vertices in the in-neigh\-bor\-hood of $v$; and the \emph{rel-1 score} of~$v$, denoted by $\mathrm{relScore}^1\!(v)$, is the number of
	\hidden{document }%
vertices in the in-neigh\-bor\-hood of $v$ whose RCC is greater than 1%
	\TechReportVersion{~(i.e., their document has already contributed to at least one solution)}%
. Similarly, the \emph{rcc-2 score} and \emph{rel-2 score} of $v$, denoted by $\mathrm{rccScore}^2\!(v)$ and $\mathrm{relScore}^2\!(v)$, respectively, focus on the 2-step in-neigh\-bor\-hood.
To define these scores formally,
\PaperVersion{%
	let $\mathrm{in}^k\!(v)$ denote the set of vertices in the $k$-step in-neigh\-bor\-hood of~$v$, and, if 
}%
\TechReportVersion{%
	let $\mathrm{in}^1\!(v)$ and $\mathrm{in}^2\!(v)$ denote the set of vertices in the 1-step and the 2-step in-neigh\-bor\-hood of~$v$, respectively; that~is,
	\begin{align*}
		\mathrm{in}^1\!(v) &= \big\lbrace v'\! \in V \,\big|\, \tuple{v'\!,v} \in E \big\rbrace \text{, and} \\
		\mathrm{in}^2\!(v) &= \mathrm{in1}(v) \cup \bigl(\, \textstyle\bigcup_{v'\! \in \mathrm{in}^1\!(v)} \mathrm{in}^1\!(v') \,\bigr) .
	\end{align*}
	If
}%
$v$ is a document vertex, let $\mathrm{rcc}(v)$ be its (current)  RCC.
Then, for each vertex $v \!\in\! V$\! and $k \in \lbrace 1,2 \rbrace$, the scoring functions are defined as~follows:
\begin{align*}
	\mathrm{rccScore}^k\!(v) &=
		\textstyle \sum_{v'\! \in \mathrm{in}^k\!(v)} \mathrm{rcc}(v')
	, \\
	\mathrm{relScore}^k\!(v) &= \bigl| \lbrace v'\! \in \mathrm{in}^k\!(v) \,|\, \mathrm{rcc}(v') > 0 \rbrace \bigr| .
\end{align*}

These vertex scoring functions can be used as a basis~for
	\removable{(local processing aware)}
graph-based approaches to prioritize URI lookups (in the same manner as the \PageRank\ and \indegree\ approaches use the PageRank algorithm and indegree-based scoring, respectively%
). Hereafter, we refer to the four resulting URI prioritization approaches
as \rccOne, \rccTwo, \relOne, and \relTwo, respectively%
\TechReportVersion{%
	~(i.e., the \rccOne\ approach uses the rcc-1 score, \rccTwo\ uses the rcc-2 score,~%
		etc.)%
}%
.

\subsection{Hybrid
Approaches} \label{ssec:Approaches-Hybrid}


	\PaperVersion{The idea of the \is\ approach can be combined with the so\-lu\-tion-aware graph-based approaches. }%
	\TechReportVersion{While both the idea of intermediate solution driven approaches and the idea of so\-lu\-tion-aware graph-based approaches can be considered separately, these ideas can also be combined. }%
To this end, the \drop\ has~to obtain via the feedback channel from the dispatcher both the provenance annotation of each solution and all intermediate solutions. Based on the provenance annotations, the \drop\ increases the RCCs of document vertices in the link graph that the operator builds incrementally (as discussed in Section~\ref{ssec:Approaches-RCCGraphBased}). The intermediate solutions are used
	to maintain an additional number for every URI \removable{that is} queued for lookup;
this number represents the maximum of the number of Covered bits set in each intermediate solution that binds some variable to the URI. Hence, initially~(i.e., when the URI is \removable{added to the lookup} queue) this number is 0, and it may increase
%
	as the \drop\ gets to see more and more intermediate solutions via the feedback channel. Observe
		that this number is always equal to the lookup priority that the \is\ approach would ascribe to the URI.
	Therefore, we call this number the \emph{IS-score} of~the~URI.

Given such IS-scores, we consider four different
	\removable{(hybrid)}
approaches to prioritize URI lookups, each of which uses one of the RCC-based vertex scoring functions introduced in Section~\ref{ssec:Approaches-RCCGraphBased}.
	We call these approaches \isrccOne, \isrccTwo, \isrelOne, and \isrelTwo~(the name indicates the vertex scoring function used). Each of them determines
the priority of a queued URI by
	multiplying the current IS-score of the URI by the current RCC-based score that their vertex scoring function returns for the URI vertex representing the URI.
%
Whenever the \drop\ 
	has to increase the IS-score of a URI
or the RCC-score of the corresponding URI vertex changes (as a result of an augmentation of the in\-cre\-men\-tal\-ly-built link graph), then the lookup priority of that URI is adapted accordingly.

\PaperVersion{
}

\section{Evaluation} \label{sec:Evaluation}

\PaperVersion{%
	This section evaluates the URI prioritization approaches introduced in the previous section. 
	We first describe the experimental setup and, thereafter, present the results of our experiments.~All digital artifacts required for our evaluation (software, data) are available
		\removableAltOn{on the Web page for the paper}{online}
	at {\small \url{http://squin.org/experiments/WWW2016/}}.

	\subsection{Experimental Results} \label{ssec:Evaluation:Results}
}

To experimentally analyze the URI prioritization approaches introduced in Section~\ref{sec:Approaches} we used each of these approaches for tra\-vers\-al-based query executions over our test Webs.
\TechReportVersion{%
	This section presents our measurements for these executions and discusses the results.
	\par
}%
The charts in Figure~\ref{fig:PrimaryMeasurements} illustrate the mean \textsf{\small relRT1st} and
\PaperVersion{%
	\hidden{the mean} \textsf{\small relRTCmpl} measured for query executions over test Web~$\symWoD_\mathsf{test}^{62,47}$\!~(in some cases the bars for \textsf{\small relRT1st} are too small to be seen). \removable{For instance, the leftmost bars in Figure~\ref{sfig:PrimaryMeasurements:Q1} indicate that, for the \baseline\ executions of query~Q1, the mean \textsf{\small relRT1st} is 0.265 and the mean \textsf{\small relRTCmpl} is 0.992. \hidden{Hence, these executions returned a first solution of the query result after 26.5\% of the overall query execution time, and it took them more than 99\% of the time to complete the query result. }}%
%
}%
\TechReportVersion{%
	\textsf{\small relRTCmpl} measured for query executions over test Web~$\symWoD_\mathsf{test}^{62,47}$\!~(in some cases the bars for \textsf{\small relRT1st} are too small to be seen).
}%
In the following, we discuss these measurements, as well as further noteworthy
	behavior as observed for query executions over the other test Webs.
The discussion is organized based on the classification of URI prioritization approaches as introduced in Section~\ref{sec:Approaches}~(cf.~Figure~\ref{fig:taxonomy}). However, we begin
	\TechReportVersion{our discussion }%
with some general observations.

\TechReportVersion{%
\begin{figure*}[t]
\centering

	\subfigure[\small Query Q1]{%
		\includegraphics[width=0.33\textwidth]{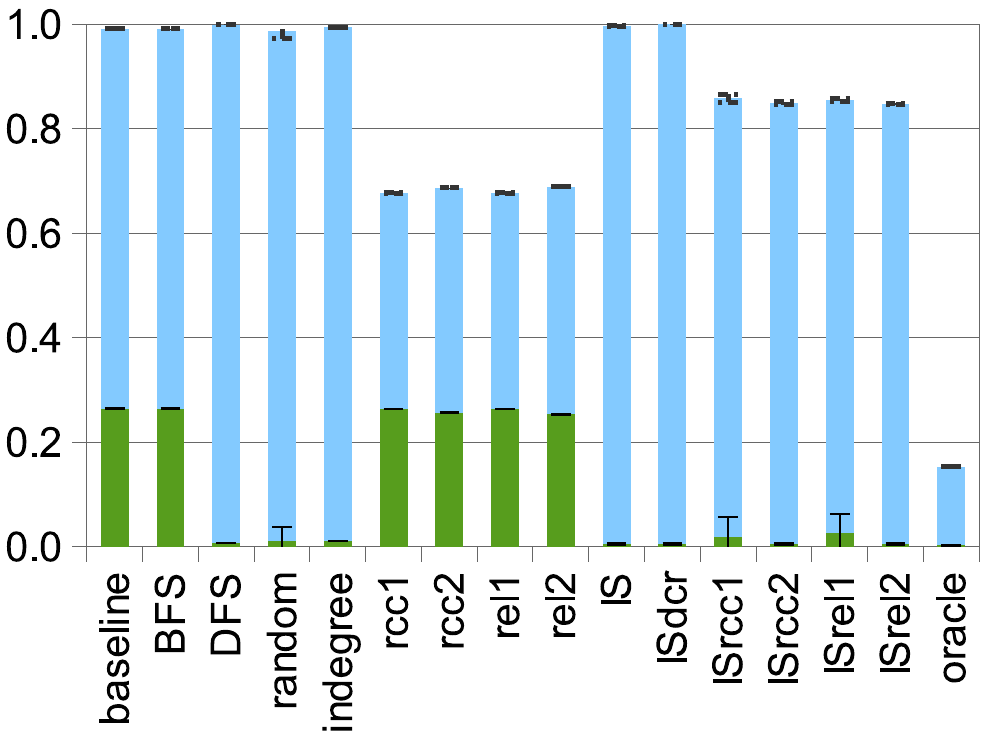}%
		\label{sfig:PrimaryMeasurements:Q1}
	}%
	\subfigure[\small Query Q2]{%
		\includegraphics[width=0.33\textwidth]{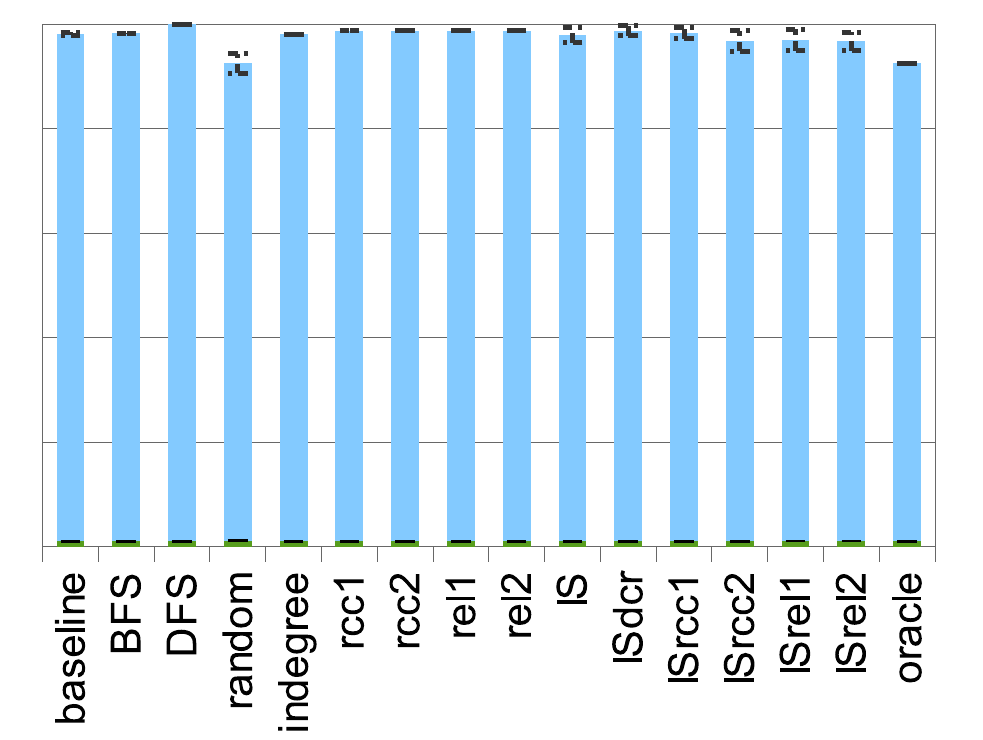}%
		\label{sfig:PrimaryMeasurements:Q2}
	}%
	\subfigure[\small Query Q3]{%
		\includegraphics[width=0.33\textwidth]{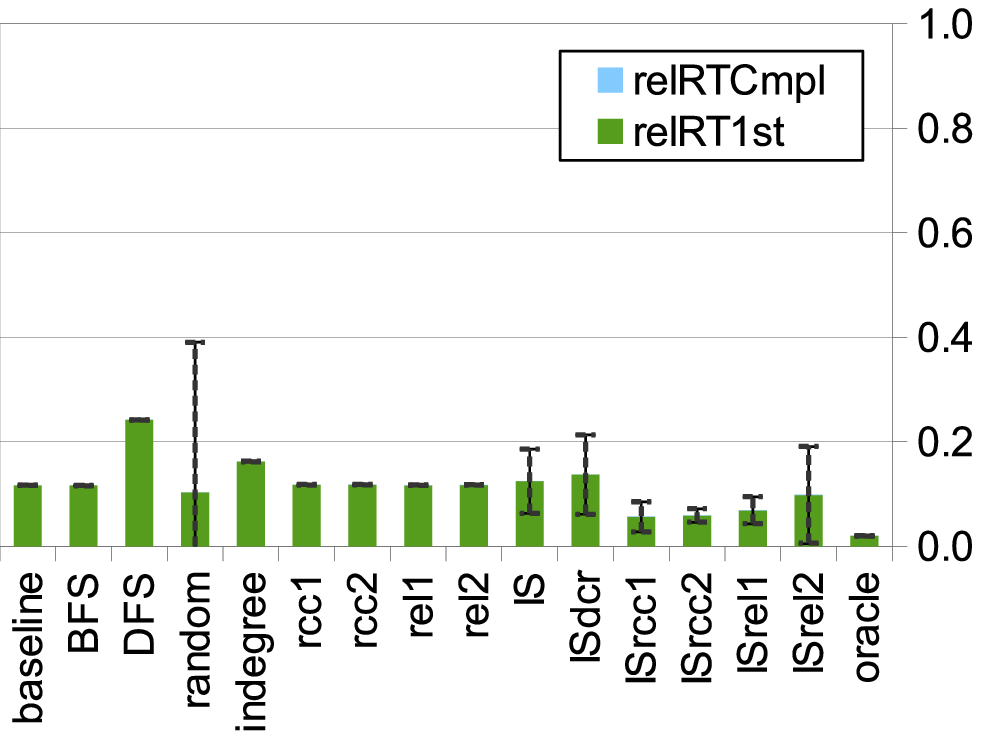}
		\label{sfig:PrimaryMeasurements:Q3}
	}

	\subfigure[\small Query Q4]{%
		\includegraphics[width=0.33\textwidth]{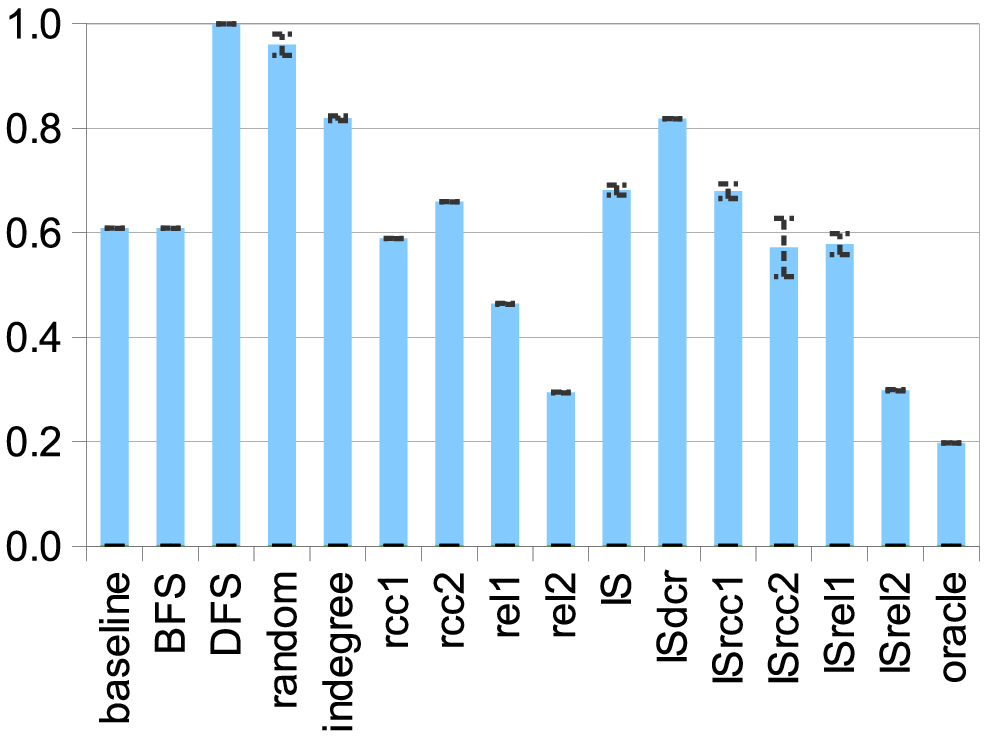}%
		\label{sfig:PrimaryMeasurements:Q4}
	}%
	\subfigure[\small Query Q5]{%
		\includegraphics[width=0.33\textwidth]{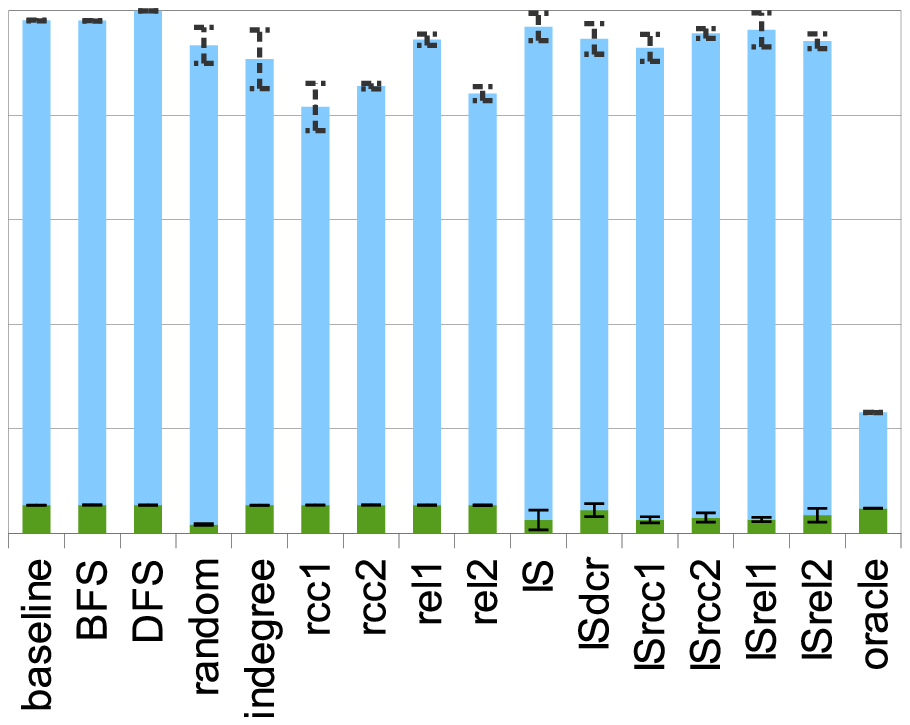}%
		\label{sfig:PrimaryMeasurements:Q5}
	}%
	\subfigure[\small Query Q6]{%
		\includegraphics[width=0.33\textwidth]{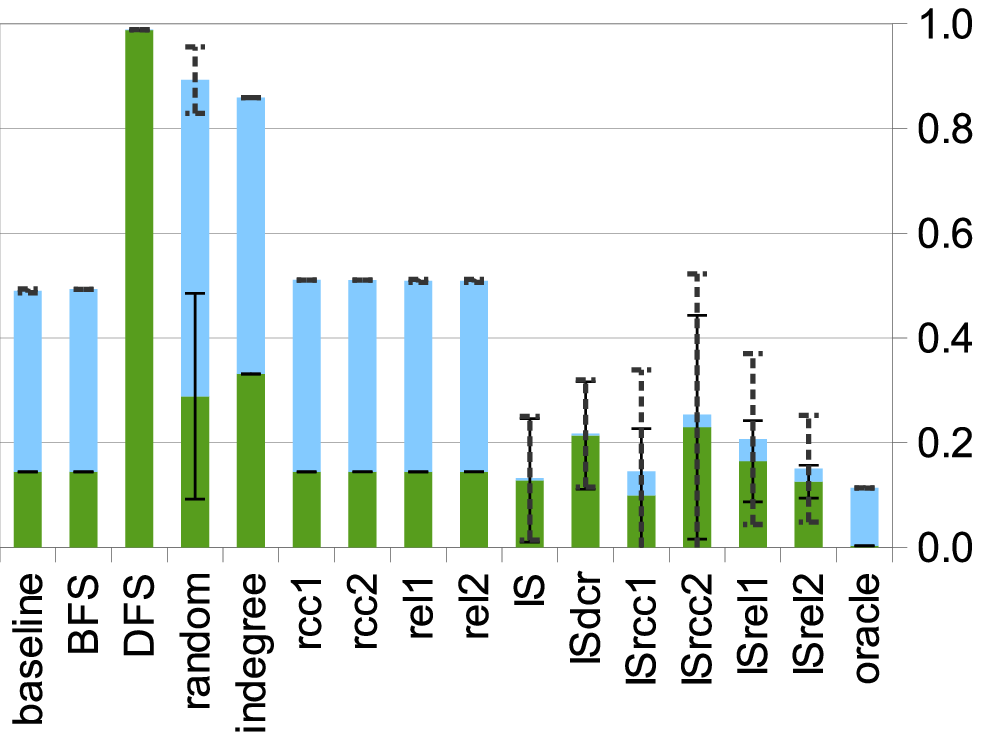}%
		\label{sfig:PrimaryMeasurements:Q6}
	}

	\caption{Relative response times
		achieved by different approaches to prioritize URI lookups (test Web: $\symWoD_\mathsf{test}^{62,47}$).}
	\label{fig:PrimaryMeasurements}
\end{figure*}
}

\PaperVersion{%
\begin{figure}[t]
\centering
	\subfigure[\small Query Q1]{%
		\includegraphics[width=0.48\columnwidth]{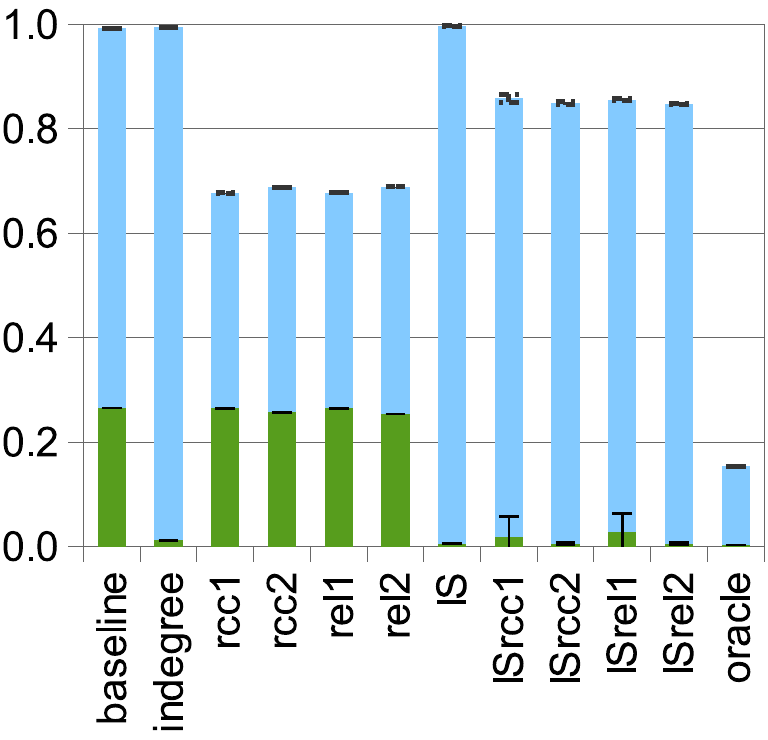}%
		\label{sfig:PrimaryMeasurements:Q1}%
	}
	\subfigure[\small Query Q2]{%
		\includegraphics[width=0.48\columnwidth]{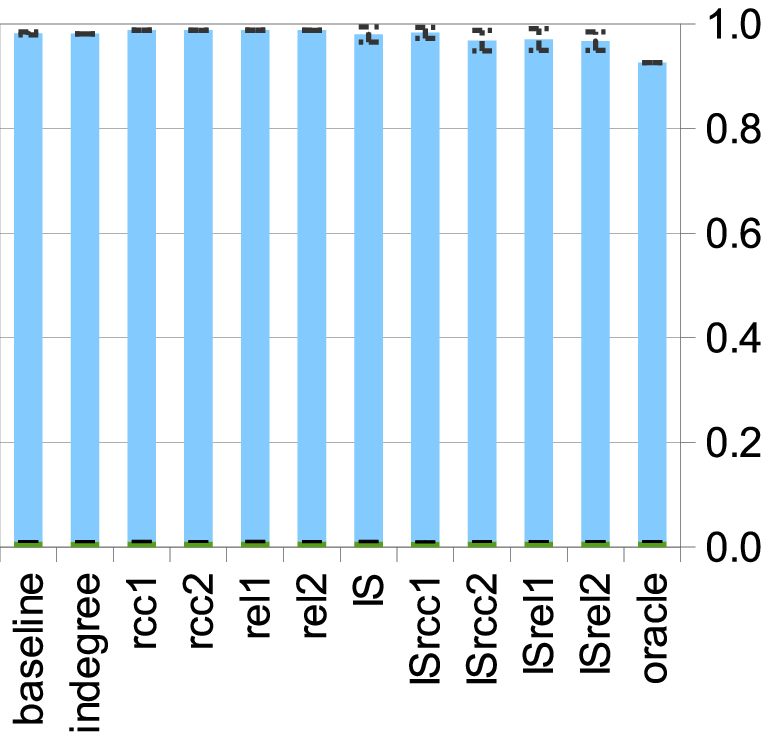}%
		\label{sfig:PrimaryMeasurements:Q2}
	}

\vspace{-2mm} 
	\subfigure[\small Query Q3]{%
		\includegraphics[width=0.48\columnwidth]{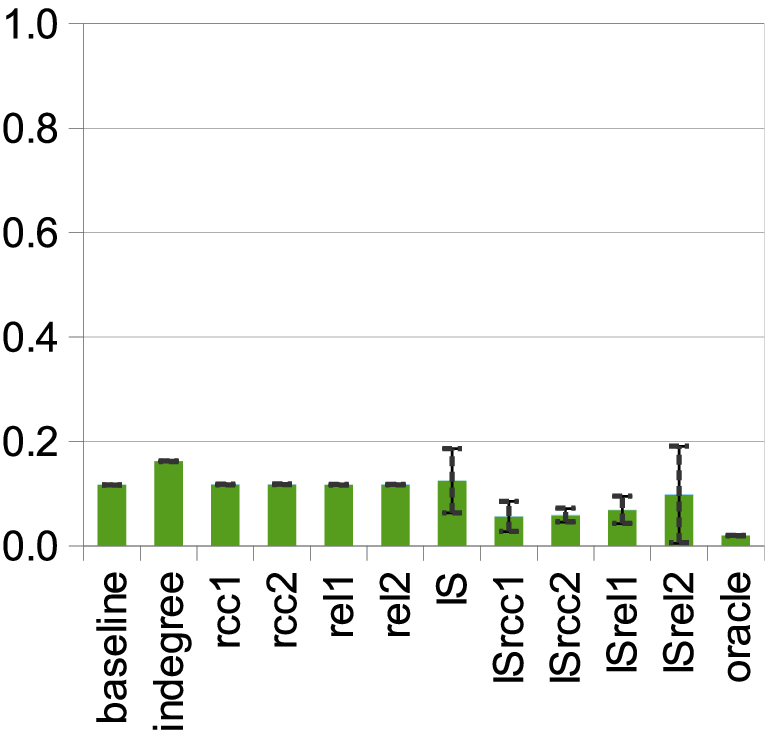}%
		\label{sfig:PrimaryMeasurements:Q3}
	}
	\subfigure[\small Query Q4]{%
		\includegraphics[width=0.48\columnwidth]{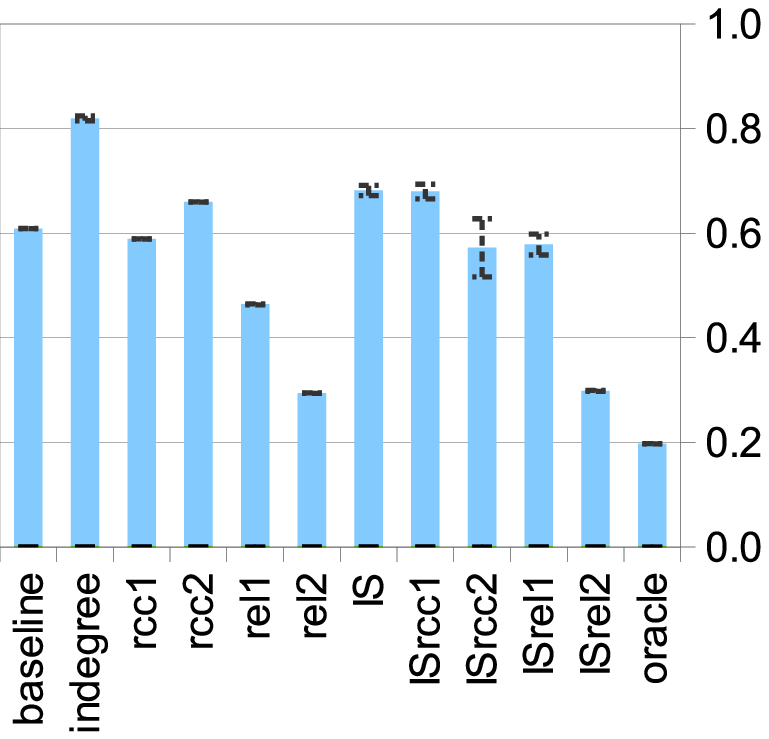}%
		\label{sfig:PrimaryMeasurements:Q4}
	}

\vspace{-2mm} 
	\subfigure[\small Query Q5]{%
		\includegraphics[width=0.48\columnwidth]{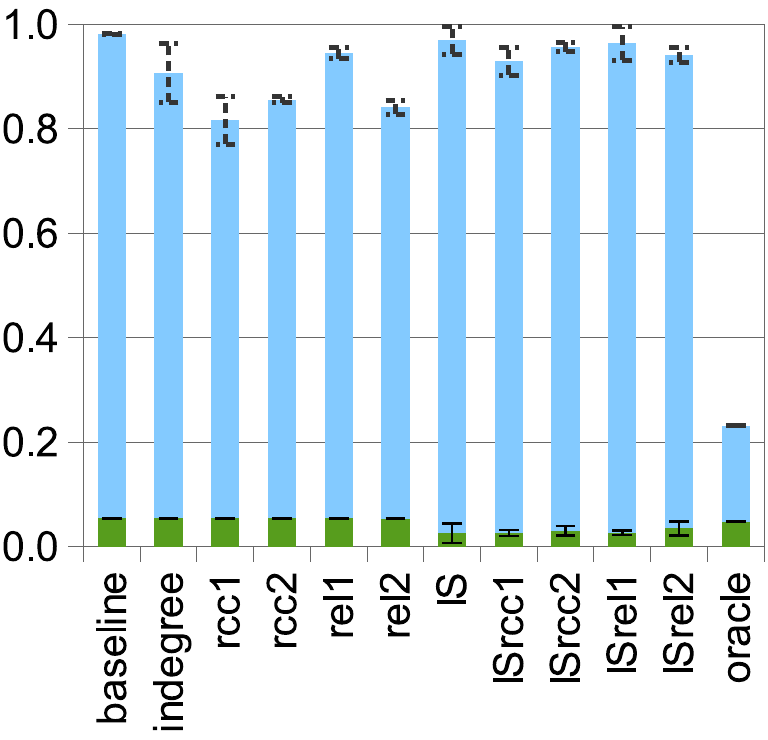}%
		\label{sfig:PrimaryMeasurements:Q5}
	}
	\subfigure[\small Query Q6]{%
		\includegraphics[width=0.48\columnwidth]{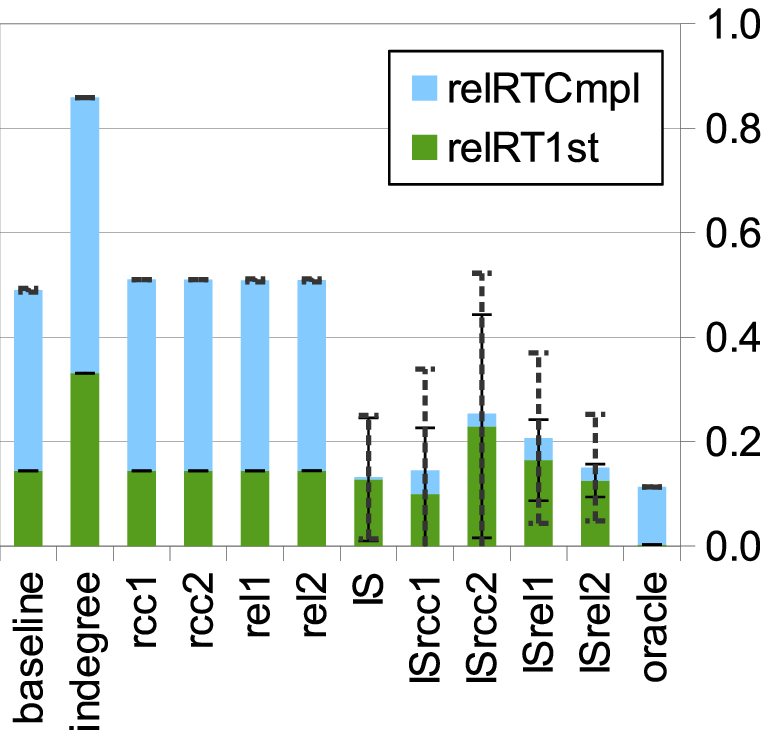}%
		\label{sfig:PrimaryMeasurements:Q6}
	}

\vspace{-3mm} 
	\caption{Relative response times
		achieved by different approaches to prioritize URI lookups (test Web: $\symWoD_\mathsf{test}^{62,47}$).}
\vspace{-2mm} 
	\label{fig:PrimaryMeasurements}
\end{figure}
}

\PaperVersion{%
	\subsubsection{General Observations}%
}%
\TechReportVersion{%
	\subsection{General Observations}%
}%
\label{sssec:Evaluation:Results:General}

\PaperVersion{%
A first, unsurprising observation is
}%
\TechReportVersion{%
As a first
observation we note that in all our experiments the \BF\ approach shows the same behavior as the \baseline\ approach. This correspondence is not surprising given that, by using a FIFO queue, the \baseline\ approach also resembles a breadth-first
	traversal.

	A similarly unsurprising observation is
}%
that, in almost all cases, none of the approaches achieves response times
	\hidden{that are }%
smaller than the response times
	\removableAltOn{achieved with}{of}
the \oracle\ approach.
%
%
However,
	\removableAltOn{we also notice}{there exist}
a few (minor) exceptions%
	\hidden{ to this general observation}%
.
	\PaperVersion{%
		\hidden{For instance, the \isrelOne-based executions of test query~Q5 over test Web~$\symWoD_\mathsf{test}^{62,47}$ achieve a mean \textsf{\small relRT1st} of 0.026 which is smaller than the 0.048 achieved by the \oracle-based executions (cf.~Figure~\ref{sfig:PrimaryMeasurements:Q5}).}%
	}%
	\TechReportVersion{%
		For instance, the \random-based executions of test query~Q5 over test Web~$\symWoD_\mathsf{test}^{62,47}$ achieve a mean \textsf{\small relRT1st} of
			0.02 which is smaller than the 0.05
		achieved by the \oracle-based executions (cf.~Figure~\ref{sfig:PrimaryMeasurements:Q5}).
	}%
These exceptions can be explained by the fact that---independent of what URI prioritization approach is applied---the \drop\ discovers the URIs to be looked up only gradually. Then,
		by \emph{greedily} ordering the currently available URIs (based on our pre-com\-put\-ed RCCs),
the \oracle\ approach may only achieve a local optimum but not the global~one.


Ignoring the \oracle\ approach for a moment,
	we notice that
approaches that achieve
	\PaperVersion{a comparably good \textsf{\small relRT1st} }%
	\TechReportVersion{comparably good first-so\-lu\-tion response times }%
for a query
	do not necessarily also achieve
	\PaperVersion{a good \textsf{\small relRTCmpl} }%
	\TechReportVersion{good com\-plete-re\-sult response times }%
for that query.
	\TechReportVersion{As an example, compare the measurements for \DF\ and for \rccOne\ in Figure~\ref{sfig:PrimaryMeasurements:Q1}.}

	Another general observation is
that,
	by using different URI prioritization approaches to execute
the same query over the same test Web, the number of intermediate solutions processed by our system can vary significantly, 
and so does the number of priority changes initiated by
the adaptive approaches. 
These variances indicate that the amount of
computation within our system can differ considerably depending on which URI prioritization approach is used. Nonetheless, in all our experiments the overall time to execute the same query over the same test Web is always almost identical for the different approaches, which again illustrates the dominance of the data retrieval fraction of query execution time (as discussed in Section~\ref{SecInText:DataRetrievalDominance}), and it also verifies the comparability of our relative measurements. The only exception is the \PageRank\ approach
	for which query execution times range from 120\% to 320\% of the execution times measured for the other approaches. Hence,
		in contrast to the additional computation required for each of the other approaches, the frequent execution of the iterative PageRank algorithm becomes a non-negligible overhead.
As a result, the \PageRank\ approach cannot compete with the other approaches and, thus, we ignore it%
	\PaperVersion{\removable{\ in the remainder of our analysis}.}%
	\TechReportVersion{ in our analysis (moreover, due to the greater query execution times, the relative response times measured for the \PageRank-based executions are not directly comparable to the relative response times measured for the other~executions).}%

Finally,
	for
query~Q3 over test Web~$\symWoD_\mathsf{test}^{62,47}$~(cf.~Figure~\ref{sfig:PrimaryMeasurements:Q3}), we note that, for
	\TechReportVersion{any approach, the mean \textsf{\small relRT1st} and the mean \textsf{\small relRTCmpl} coincide. That is, the differences between the times needed to return a first solution of the query result and the times to return the last solution are insignificant (however, these times are different for the different approaches). }%
	\PaperVersion{all approaches, the differences between the time needed to return a first solution \hidden{of the query result }and the time to return the last solution are insignificant. }%
We explain this phenomenon as follows: Only three of the 234 reachable documents for \hidden{query~}Q3 over $\symWoD_\mathsf{test}^{62,47}$~contribute to the query result and this result consists of four solutions\TechReportVersion{~(cf.~Table~\ref{tab:W-62-47})}. It turns out that the computation of each of these \TechReportVersion{four }solutions requires data from each of the \removable{three} re\-sult-rel\-e\-vant documents. Hence,
	only after (and as soon as)
the last of these three documents has been retrieved, the system can compute and return all four solutions.

\TechReportVersion{
	The remainder of this section discusses our measurements for each of the URI prioritization approaches in detail.
We begin with the non-adap\-tive approaches~(\BF, \DF, \random).
}

%

\TechReportVersion{%

\PaperVersion{%
	\subsubsection{Non-Adaptive Approaches}%
}%
\TechReportVersion{%
	\subsection{Non-Adaptive Approaches}%
}%
\label{sssec:Evaluation:Results:NonAdaptive}

\PaperVersion{%
	Due to space limitations, we only highlight our main findings for the non-adap\-tive approaches\removable{~(\BF, \DF, \random)}. For detailed discussion of these finding, refer to
		our technical report~\cite{ExtendedVersion}.

	Regarding \BF, we identify
	two characteristics of
	reachable subwebs that are beneficial for \BF:
	First, the greater the number of re\-sult-ir\-rel\-e\-vant documents that are farther from a seed document than the farthest re\-sult-rel\-e\-vant document, the lower the \textsf{\small relRTCmpl} that can be achieved by using the \BF\ approach.
	Second, the greater the number of re\-sult-rel\-e\-vant documents that are closer to a seed document than the majority of re\-sult-ir\-rel\-e\-vant documents, the higher the chance that the \BF\ approach achieves a low \textsf{\small relRT1st}.
} 

\TechReportVersion{%
Given the measurements in Figure~\ref{fig:PrimaryMeasurements}, we observe that the \BF\ approach is more effective in achieving low first-so\-lu\-tion response times for queries Q2, Q4, and Q5 than it is for Q1, Q3, and Q6. If we consider com\-plete-re\-sult response times on the other hand, the \BF\ approach is effective only for Q3, Q4, and~Q6.

We explain these observations based on the statistics in Table~\ref{tab:W-62-47}. The ``max'' columns in this table indicate that, for queries Q3-Q6,
the
	\removable{que\-ry-spe\-cif\-ic}
reachable subwebs of $\symWoD_\mathsf{test}^{62,47}$ contain some re\-sult-ir\-rel\-e\-vant documents that are farther away from some seed document than the farthest re\-sult-rel\-e\-vant documents%
	\removable{~(e.g., for Q6, some irrelevant documents are as much as four hops farther than the farthest relevant document)}%
. The \BF\ approach discovers and retrieves these irrelevant documents only after retrieving all re\-sult-rel\-e\-vant documents.
	At this point, the execution can already compute and return all solutions of the corresponding query result. While this explains why the \BF\ executions of queries Q3, Q4, and Q6
over $\symWoD_\mathsf{test}^{62,47}$ achieve \textsf{\small relRTCmpl} values that are significantly lower than the \textsf{\small relRTCmpl} measured for Q1 and Q2, the high \textsf{\small relRTCmpl} values for Q5 are somewhat surprising. However, a closer inspection of the Q5-spe\-cif\-ic reachable subweb of $\symWoD_\mathsf{test}^{62,47}$ reveals that, in this subweb, it is only a single irrelevant document that is farther from the seed than the farthest relevant documents. This one document is not enough to make a significant difference to reachable subwebs in which the farthest irrelevant documents are as far from a seed as the farthest relevant documents (as is the case in the Q1-spe\-cif\-ic and the Q2-spe\-cif\-ic reachable subweb of $\symWoD_\mathsf{test}^{62,47}$\!, respectively). Hence, we identify the following dependency:
\Finding{The greater the number of
	\removable{re\-sult}-ir\-rel\-e\-vant
documents that are farther from a seed
	\removable{document}
than the farthest
	\removable{re\-sult}-rel\-e\-vant
document, the lower the relative com\-plete-re\-sult response times that can be achieved by using the \BF\ approach.}

In a similar manner we may explain the aforementioned observation regarding the \textsf{\small relRT1st} values achieved by the \BF\ executions. Due to space constraints, we omit this discussion and only mention the following dependency that we identify
	by analyzing \textsf{\small relRT1st} values of \BF\ executions in our test Webs
(the ``mean'' and ``min'' columns in Table~\ref{tab:W-62-47} are a preliminary indication of this dependency):
\Finding{The greater the number of re\-sult-rel\-e\-vant documents that are closer to a seed document than the majority of re\-sult-ir\-rel\-e\-vant documents, the higher the chance that the \BF\ approach achieves a low
	relative first-so\-lu\-tion response time.}
} 

\PaperVersion{%
Regarding \DF, we note that, typically, the \textsf{\small relRTCmpl} values measured for \DF\ executions are very high. On the other hand, in some cases, the \DF\ approach may achieve a comparably low \textsf{\small relRT1st}; e.g., Figure~\ref{sfig:PrimaryMeasurements:Q1}. However, in the majority of cases, the \DF\ approach is not better (and often worse) than the \baseline\ and, thus, it is unsuitable for response time optimization.
} 

\TechReportVersion{%
Regarding
	\TechReportVersion{the \DF approach, we note that, for all queries over all test Webs, this approach is almost always among the worst w.r.t.~com\-plete-re\-sult response times; i.e., }%
	\PaperVersion{\DF, we note that, }%
typically, the \textsf{\small relRTCmpl} values measured for \DF\ executions are very high. This behavior can be attributed to the fact that the re\-sult-rel\-e\-vant documents in a reachable subweb are unlikely to be found along only one of the paths that the \DF\ approach traverses one after another. Instead, in most reachable subwebs multiple of these depth-first traversal paths contain re\-sult-rel\-e\-vant documents (the Q3-spe\-cif\-ic reachable subweb of $\symWoD_\mathsf{test}^{62,47}$ is one of the few exceptions). Then, in many cases, the \DF\ approach traverses some of these paths with re\-sult-rel\-e\-vant documents only close to the end of the overall traversal process, which results in a com\-plete-re\-sult response time that is close to the overall query execution time.
	In the worst case, all of the paths with relevant documents are traversed at the end, which has a negative impact not only on the com\-plete-re\-sult response time, but also on the first-so\-lu\-tion response time~(see Figure~\ref{sfig:PrimaryMeasurements:Q6} for an example).
On the other hand, it is possible
	that, during an early stage of the traversal process, the \DF\ approach traverses a path that
contains every (relevant) document necessary for completing the computation of a first query solution. In such a case, the \DF\ approach may achieve a comparably low first-so\-lu\-tion response time~(e.g., Figure~\ref{sfig:PrimaryMeasurements:Q1}). 
However, in the majority of cases, the \DF\ approach is not better (and often worse) than the \baseline\ and \BF\ approaches and, thus, it is unsuitable for response time optimization.
} 

Similarly, our measurements show that the \random\ approach also is unsuitable. In particular, \random\ achieves
	\PaperVersion{\textsf{\small relRTCmpl} values }%
	\TechReportVersion{com\-plete-re\-sult response times }%
that are often worse than the baseline. For
	\PaperVersion{\textsf{\small relRT1st}, }%
	\TechReportVersion{first-so\-lu\-tion response time, }%
in most cases \random\ is similar to \baseline. Another disadvantage of the \random\ approach is its unpredictability; that is, measurements for the same query over the same test Web can vary significantly.
For instance, the error bars
in the charts in
	Figures \ref{sfig:PrimaryMeasurements:Q3} and~\ref{sfig:PrimaryMeasurements:Q6}
highlight these variations for the \random-based executions of queries Q3 and Q6%
	\TechReportVersion{ over test Web $\symWoD_\mathsf{test}^{62,47}$\!}%
	.


} 

\PaperVersion{%
	\subsubsection{Purely Graph-Based Approaches}%
}%
\TechReportVersion{%
	\subsection{Purely Graph-Based Approaches}%
}%
\label{sssec:Evaluation:Results:PurelyGraph}

After ruling out the \PageRank\ approach~(cf.~Section~\ref{sssec:Evaluation:Results:General}), \indegree\ is the only remaining purely graph-based approach
	\hidden{tested }%
in our experiments.
We observe that
	\removableAltOn{the \indegree\ }{this}
approach is often worse and only in a few cases better than the \baseline\ approach~(for both \textsf{\small relRT1st} and \textsf{\small relRTCmpl}). The reason for the negative performance of
	\removableAltOn{the \indegree\ }{this}	
approach---as well as any other possible purely graph-based approach---is that the applied vertex scoring method rates document and URI vertices only based on graph-spe\-cif\-ic properties, whereas the re\-sult-rel\-e\-vance of reachable documents is independent of such properties. In fact, in earlier work we show
	\hidden{empirically }%
that there does not exist a correlation between the indegree (or the PageRank, the HITS scores, the $k$-step Markov score, or the betweenness centrality) of document vertices in the link graph of reachable subwebs and the re\-sult-(ir)rel\-e\-vance of the corresponding documents~\cite{Hartig14:ReachableSubwebs}.

\PaperVersion{%
	\subsubsection{Solution-Aware Graph-Based Approaches}%
}%
\TechReportVersion{%
	\subsection{Solution-Aware Graph-Based Approaches}%
}%
\label{sssec:Evaluation:Results:SolAware}

In contrast to the purely graph-based approaches,
	\TechReportVersion{the }%
so\-lu\-tion-aware graph-based approaches%
	\TechReportVersion{, \rccOne, \rccTwo, \relOne, and \relTwo, } employ vertex scoring methods that make use of information about re\-sult-rel\-e\-vant documents as discovered
during the query execution process. We notice that until such information becomes available~(%
	\PaperVersion{i.e., }%
	\TechReportVersion{that is, }%
not before a first query solution has been computed), these methods rate all vertices equal. As a consequence, all URIs added to the lookup queue have the same priority and are processed in the order in which they are discovered. Hence, until a first solution has been computed, the so\-lu\-tion-aware graph-based approaches behave like the \baseline\ approach. Therefore, these approaches always achieve the same
	\PaperVersion{\textsf{\small relRT1st} }%
	\TechReportVersion{first-so\-lu\-tion response time }%
as the \baseline.

Once a first set of re\-sult-rel\-e\-vant documents can be identified, the so\-lu\-tion-aware graph-based approaches begin leveraging this information. As a result, for several query executions in our experiments, these approaches achieve
	\PaperVersion{a \textsf{\small relRTCmpl} that is }%
	\TechReportVersion{com\-plete-re\-sult response times that are }%
significantly lower than the baseline. Moreover, for the majority of query executions for which this is not the case, the
	\PaperVersion{\textsf{\small relRTCmpl} achieved by the so\-lu\-tion-aware graph-based approaches is }%
	\TechReportVersion{com\-plete-re\-sult response times achieved by the so\-lu\-tion-aware graph-based approaches are }%
comparable to the baseline.
%
In the following, we identify characteristics of reachable subwebs that are beneficial for our four so\-lu\-tion-aware graph-based
	\PaperVersion{approaches (for a more detailed discussion refer to~\cite{ExtendedVersion}).}%
	\TechReportVersion{approaches and compare these approaches with each other.}

\PaperVersion{%
	A necessary
		\hidden{(but not necessarily sufficient) }%
	characteristic is that every
		reachable document that is re\-sult-rel\-e\-vant
	must have at least one in-neigh\-bor that is also re\-sult-rel\-e\-vant~(for \relTwo\ and \rccTwo\ it may also be a 2-step in-neigh\-bor).
	However, even if the in-neigh\-bor\-hood of a relevant document $d$ contains some other relevant documents, the so\-lu\-tion-aware graph-based approaches can increase the retrieval priority of document $d$ only if the relevance of at least one of these other documents, say $d'$\!, is discovered before the retrieval of $d$. This is possible only if the relevance of $d'$ can be attributed to its contribution to some query solution whose computation does not require document~$d$.
} 

\TechReportVersion{%
A first, necessary
	\removable{(but not necessarily sufficient)}
characteristic is that
\Finding{%
every
	reachable document that is re\-sult-rel\-e\-vant
%
	must have
at least one in-neigh\-bor
	that
is also re\-sult-rel\-e\-vant%
}~%
(for \relTwo\ and \rccTwo\ it may also be a 2-step in-neigh\-bor).
	Then, the document can be assigned a higher retrieval priority when the relevance of its in-neigh\-bor has been identified. Hence, for reachable subwebs without this characteristic,
the so\-lu\-tion-aware graph-based approaches are unlikely to achieve an advantage over the \baseline\ approach.
\TechReportVersion{
	For instance, the Q2-spe\-cif\-ic reachable subweb of test Web $\symWoD_\mathsf{test}^{62,47}$ contains two re\-sult-rel\-e\-vant documents whose in-neigh\-bor\-hoods consist of irrelevant documents only. Therefore, the so\-lu\-tion-aware graph-based approaches can never assign the retrieval of these two documents a high priority whenever information about other relevant documents becomes available. Since these two documents are comparably far from a seed document (7 and 8 hops, respectively), they are thus retrieved close to the end of the corresponding query executions. As a consequence, the \textsf{\small relRTCmpl} values of these executions are high~(cf.~Figure~\ref{sfig:PrimaryMeasurements:Q2}).
}

However, even if the in-neigh\-bor\-hood of a relevant document $d$ contains some other relevant documents, the so\-lu\-tion-aware graph-based approaches can increase the retrieval priority of document $d$ only if the relevance of at least one of these other documents, say $d'$\!, is discovered before the retrieval of $d$. This is possible only if the relevance of $d'$ can be attributed to its contribution to some query solution whose computation does not require document $d$. Then, if \removable{$d'$ and all other documents required for computing this solution have been retrieved before the retrieval of $d$ (and, thus,} the solution can be computed before retrieving $d$), the so\-lu\-tion-aware graph-based approaches can increase the priority for $d$.
} 

\PaperVersion{%
	As can be seen from this discussion, an early computation of some solutions, resulting from an early retrieval of some re\-sult-rel\-e\-vant documents, may increase chances that the so\-lu\-tion-aware graph-based approaches retrieve all relevant documents early and, thus, may improve com\-plete-re\-sult response times. However, there are also cases in which the identification of relevant documents may mislead these approaches; in particular, if some relevant documents link to many irrelevant documents.
	A special case that is particularly worse for \rccOne\ and \rccTwo\ is the existence of a re\-sult-rel\-e\-vant document with an RCC that is significantly higher than the RCCs of the other relevant documents in this subweb; such a high RCC may dominate the RCC-based scores in the in-neighborhood of the document. The Q4-spe\-cif\-ic reachable subweb of
		\removable{test Web}
	$\symWoD_\mathsf{test}^{62,47}$ is an example of such a case~(cf.~Figure~\ref{sfig:PrimaryMeasurements:Q4}).
} 

\TechReportVersion{%
As can be seen from this discussion, an early computation of some solutions, resulting from an early retrieval of some re\-sult-rel\-e\-vant documents, may increase chances that the so\-lu\-tion-aware graph-based approaches retrieve all relevant documents early and, thus, may improve com\-plete-re\-sult response times. However, there are also cases in which the identification of relevant documents may mislead these approaches; in particular, if some relevant documents link to many irrelevant documents, these irrelevant documents also receive a high retrieval priority as a result of identifying relevant documents in their in-neigh\-bor\-hood.
\PaperVersion{%
	An extreme example of such a case is the Q4-spe\-cif\-ic reachable subweb of $\symWoD_\mathsf{test}^{62,47}$~(cf.~Figure~\ref{sfig:PrimaryMeasurements:Q4}), which contains a re\-sult-rel\-e\-vant document with an RCC that is significantly higher than the RCCs of the other relevant documents in this subweb. When the first solutions based on data from this document have been computed, the high RCC starts dominating the RCC-based scores in the neighborhood of the document, which includes many irrelevant documents. For the rcc-2 score, this dominance affects not only the 1-step in-neighborhood, but also the 2-step in-neighborhood. 
	The \relOne\ and \relTwo\ approaches are not affected by this particular issue because rel-1 and rel-2 scores cannot be dominated by documents with an unusually high RCC.
}%
\TechReportVersion{%
	An extreme example of such a case is the Q4-spe\-cif\-ic reachable subweb of $\symWoD_\mathsf{test}^{62,47}$\!. Figure~\ref{fig:ExampleOutdegreeVsRCC} illustrates the RCC and the outdegree of all documents in this subweb. Observe the existence of a re\-sult-rel\-e\-vant document that has
		an RCC of 631, \removable{which is significantly higher than the RCCs of the other relevant documents}.
	This document links to~40 other documents, many of which are re\-sult-ir\-rel\-e\-vant. Moreover, the document is only one hop from the seed document and, thus, the so\-lu\-tion-aware graph-based approaches retrieve it early. After its retrieval and the retrieval of some of the few relevant documents linked from it, the query execution starts computing the first of the 631 solutions contributed to by the document. At this point, the growing RCC of the document starts dominating the RCC-based scores in its neighborhood. For the rcc-2 score, this dominance affects not only the documents one hop away, but also those that are two hops from the dominating document. As a consequence, the \rccTwo\ approach is misled to assign a high retrieval priority to a large number of re\-sult-ir\-rel\-e\-vant documents, which has a negative impact on the com\-plete-re\-sult response time achieved by the approach~(cf.~Figure~\ref{sfig:PrimaryMeasurements:Q4}). The \relOne\ and \relTwo\ approaches are not affected by this issue because rel-1 and rel-2 scores cannot be dominated by documents with an unusually high RCC.

\begin{figure}[t]
	\centering
	\includegraphics[width=\columnwidth]{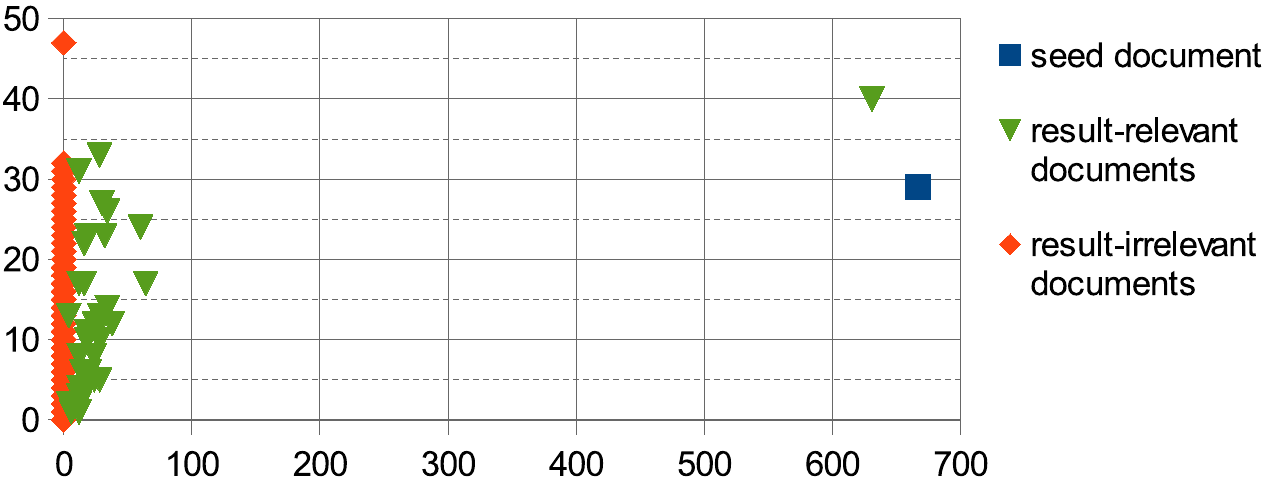}
	\caption{RCC (x-axis) vs.~outdegree (y-axis) of all documents in the Q4-spe\-cif\-ic reachable subweb of test Web $\symWoD_\mathsf{test}^{62,47}$\!.}
	\label{fig:ExampleOutdegreeVsRCC}
\end{figure}
}

} 

\PaperVersion{%
	\subsubsection{Intermediate Solution Driven Approaches}%
}%
\TechReportVersion{%
	\subsection{Intermediate Solution Driven Approaches}%
}%
\label{sssec:Evaluation:Results:ISDriven}

Intermediate solution driven approaches (including the hybrid approaches analyzed in the following section) use information about every intermediate solution generated by the
	\tpops.
However, what intermediate solutions are generated exactly depends on the routing policy applied by the dispatcher. As a consequence, the behavior of any intermediate solution driven approach depends on the routing policy.
\TechReportVersion{%
	To study whether this dependency has an impact on the response times achieved by intermediate solution driven approaches and to select a routing policy for the evaluation in this paper, we conducted an initial experiment. In the following, we discuss the results of the experiment and, thereafter, compare the intermediate solution driven approaches to the baseline.

	For any of our intermediate solution driven approaches (including the hybrid approaches) in combination with any of the eight con\-tent-based routing policies introduced in Section~\ref{ssec:RoutingPolicies}, we executed our six test queries five times over test Web~$\symWoD_\mathsf{test}^{62,47}$\!. As an example of the measurements obtained by this experiment, Figure~\ref{fig:RoutingPolicies} illustrates the mean \textsf{\small relRT1st} and the mean \textsf{\small relRTCmpl} of the \is-based and the \isDecr-based executions of queries Q1, Q4, and Q6. The experiment shows that, for each intermediate solution driven approach, there exist a significant number of cases in which the routing policy has a considerable impact on the response times achieved by the approach (both for \textsf{\small relRT1st} and \textsf{\small relRTCmpl}).

	\begin{figure*}[t]
\centering
	\subfigure[\small Query Q1]{%
		\includegraphics[width=0.33\textwidth]{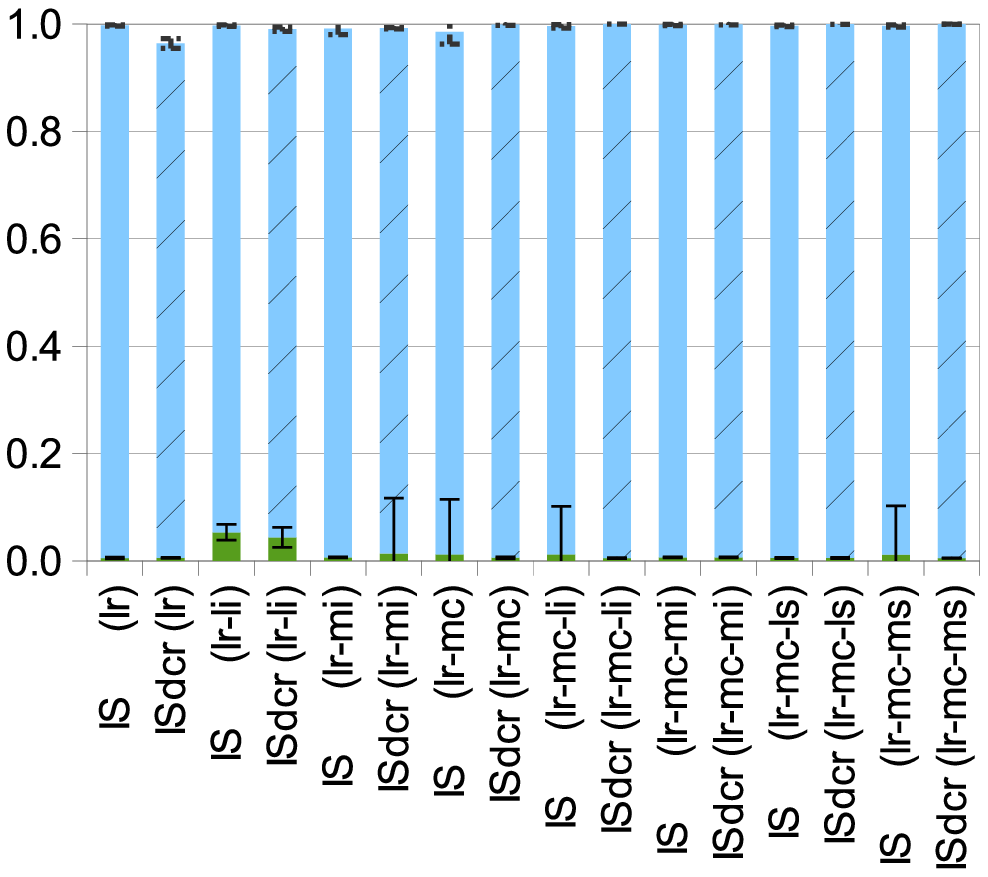}%
		\label{sfig:RoutingPolicies:Q1}
	}%
	\subfigure[\small Query Q4]{%
		\includegraphics[width=0.33\textwidth]{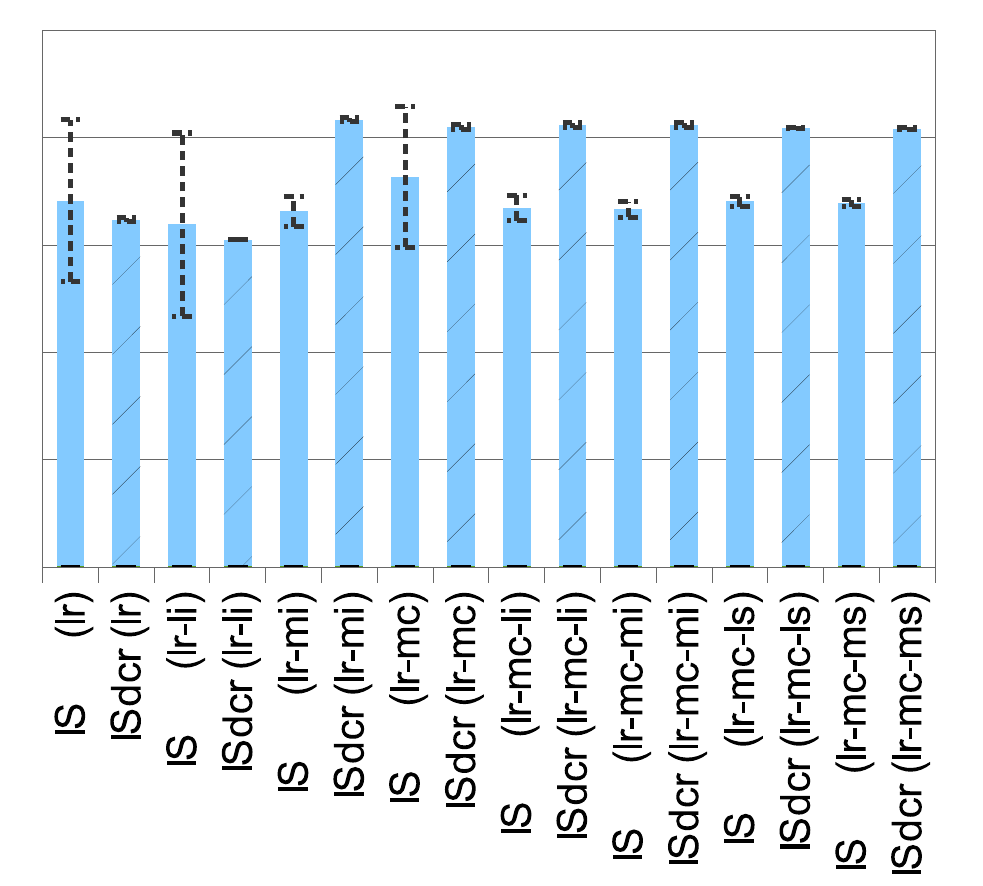}%
		\label{sfig:RoutingPolicies:Q4}
	}%
	\subfigure[\small Query Q6]{%
		\includegraphics[width=0.33\textwidth]{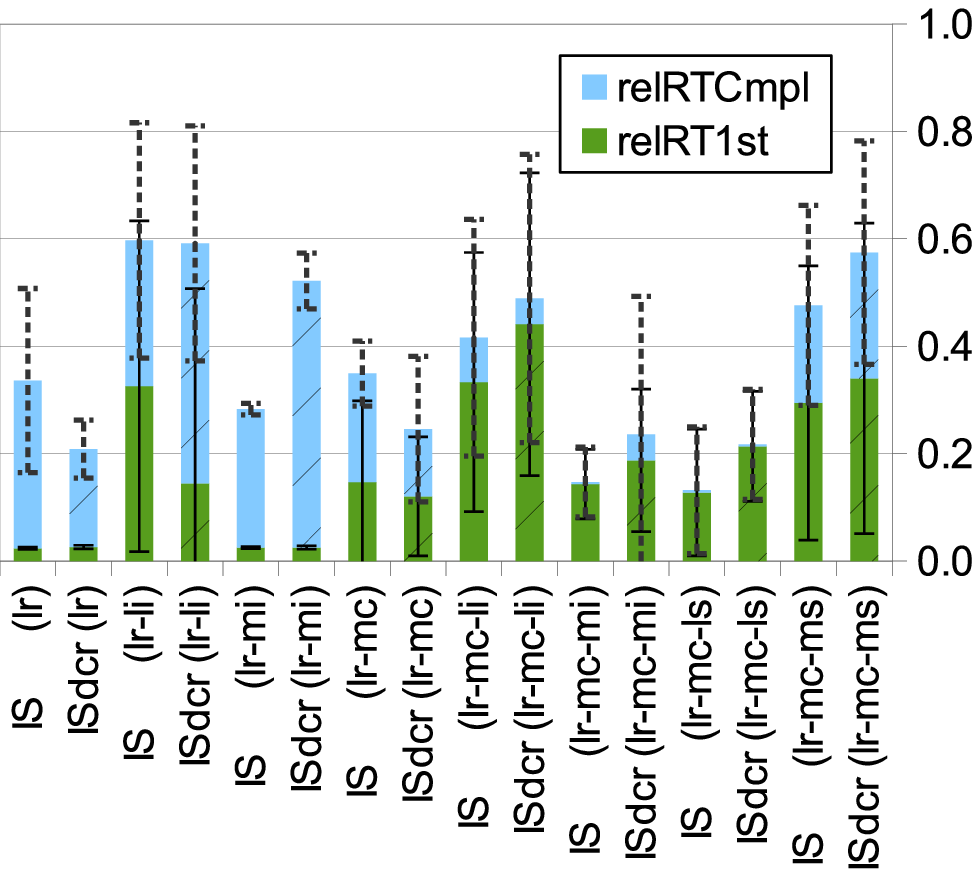}
		\label{sfig:RoutingPolicies:Q6}
	}

	\caption{Relative response times achieved by \is\ and \isDecr\ with different routing policies (test Web: $\symWoD_\mathsf{test}^{62,47}$).}
	\label{fig:RoutingPolicies}
\end{figure*}

	Moreover, for many of the five query executions, we observe a high variation in our measurements (as indicated by the error bars that represent one standard deviation in Figure~\ref{fig:RoutingPolicies}). Partially, this variation can be attributed to the randomness in the decisions of the routing policies. However, another, orthogonal reason is
}%
the
	multithreaded execution of all operators in our implementation \removable{of tra\-vers\-al-based query execution (cf.~Section~\ref{sec:Implementation})}.
Due to multithreading, the order in which intermediate solutions from different \tpops\ appear in the input queue of the dispatcher is nondeterministic, and so is the order in which the \drop\ receives information about the intermediate solutions via the feedback channel from the dispatcher. As a result, the intermediate solution driven adaptation of the priorities of URIs queued for lookup becomes nondeterministic and, thus, the order in which reachable documents are retrieved may differ for different executions of the same query. Such differences may cause different response times because the retrieval order of documents determines which intermediate solutions can be generated at which point during the query execution process. Additionally, given that the dispatcher receives and processes incoming intermediate solutions in a nondeterministic order, the order in which the \tpops\ receive their incoming intermediate solutions also becomes nondeterministic, which is another factor contributing to a diverging behavior.
\PaperVersion{%

	Irrespective of the variations, our measurements indicate that, in a number of cases, the \is\ approach
}%
\TechReportVersion{%
	Consequently, even routing policies that do not make decisions based on random cannot mitigate the observed variations.

		Considering all the routing policies studied
	in this initial experiment,
		\policy{lr-mi}
	appears to provide the best trade-off of being beneficial (w.r.t.~response times) for all intermediate solution driven approaches studied in this paper~(including the hybrid approaches). Additionally, this policy is among the policies that cause the smallest variations on average. Therefore, we selected
		\policy{lr-mi}
	as the routing policy used for
		the evaluation in
	this paper.

	Our experiments for this evaluation
		\removable{(such as the query executions whose relative response times are illustrated in Figure~\ref{fig:PrimaryMeasurements})}
	indicate that, in a number of cases, the two non-hy\-brid intermediate solution driven approaches \is\ and \isDecr\ 
}%
can achieve an advantage over the \baseline\ approach. For instance, compare the \textsf{\small relRT1st} values in Figure~\ref{sfig:PrimaryMeasurements:Q1} or the \textsf{\small relRTCmpl} values in Figure~\ref{sfig:PrimaryMeasurements:Q6}. However, there also exist a significant number of cases in which
	\PaperVersion{\is\ performs worse than the \baseline\ approach; e.g., 	Figure~\ref{sfig:PrimaryMeasurements:Q4}.}%
	\TechReportVersion{both \is\ and \isDecr\ perform worse than the \baseline\ approach (e.g., query Q4 over test Web $\symWoD_\mathsf{test}^{62,47}$\!; cf.~Figure~\ref{sfig:PrimaryMeasurements:Q4}).}%

\TechReportVersion{%
By comparing the measurements for both the \is-based
	\removable{executions}
and the \isDecr-based executions to each other, we note that both approaches often achieve similar response times.
}

\PaperVersion{%
	\subsubsection{Hybrid Approaches}%
}%
\TechReportVersion{%
	\subsection{Hybrid Approaches}%
}%
\label{sssec:Evaluation:Results:Hybrid}

For the hybrid approaches
	\TechReportVersion{(\isrccOne, \isrccTwo, \isrelOne, and \isrelTwo) }%
we first notice that these
	\TechReportVersion{also }%
achieve similar response times in many cases.
Perhaps more importantly, these response times are comparable, or at least close, to the best of either the response times achieved by the so\-lu\-tion-aware graph-based approaches
	\TechReportVersion{(\rccOne, \rccTwo, \relOne, and \relTwo) }%
or the response times of the \is-based executions.

A typical example are the executions of query Q1 over test Web $\symWoD_\mathsf{test}^{62,47}$~(cf.~Figure~\ref{sfig:PrimaryMeasurements:Q1}). On one hand, the hybrid approaches achieve
	\PaperVersion{a \textsf{\small relRTCmpl} for this query that is}
	\TechReportVersion{com\-plete-re\-sult response times for this query that are }%
small\-er than the baseline, which is also the case for the so\-lu\-tion-aware graph-based approaches but not for the \is-based executions. On the other hand, instead of also achieving
	\PaperVersion{a \textsf{\small relRT1st} as achieved by the so\-lu\-tion-aware graph-based approaches (which is as high as the baseline), the hybrid approaches achieve a \textsf{\small relRT1st} that is }%
	\TechReportVersion{the first-so\-lu\-tion response times as achieved by the so\-lu\-tion-aware graph-based approaches (which are as high as the baseline), the hybrid approaches achieve first-so\-lu\-tion response times that are }%
as small as what the \is-based executions achieve.
Regarding
	\PaperVersion{\textsf{\small relRT1st} }%
	\TechReportVersion{first-so\-lu\-tion response times }%
we recall that the so\-lu\-tion-aware graph-based approaches cannot be better than the baseline. The latter observation shows that this is not the case for the hybrid approaches. On the contrary, even if each hybrid approach is based on a so\-lu\-tion-aware graph-based approach, the combination with intermediate solution driven functionality enables the hybrid approaches to outperform the baseline w.r.t.~%
	\PaperVersion{\textsf{\small relRT1st}.}%
	\TechReportVersion{first-so\-lu\-tion response times.}



\PaperVersion{%
	\subsubsection{Comparison}%
}%
\TechReportVersion{%
	\subsection{Comparison}%
}%

Our measurements show that there is no clear winner among the URI prioritization approaches studied in this paper.
	Instead,
for each approach, there exist cases in which the approach is better than the baseline and cases in which the approach is worse%
	\hidden{~than the baseline}%
.
%
%

	Table~\ref{tab:Comparison} quantifies these cases; that is, the table
%
	lists
the percentage of cases in which the response times achieved by each approach%
	\TechReportVersion{\footnote{The \BF\ approach is not included in the table because it achieves the same response times as the baseline (cf.~Section~\ref{sssec:Evaluation:Results:General}).}}
are at least 10\% worse (resp.~10\% better) than the baseline. For this comparison, we consider the executions of all six test queries \emph{over all 14 test Webs}~(i.e., 84 cases for each approach), and we use the threshold of 10\% to focus only on noteworthy differences to the baseline.
In addition to \textsf{\small relRT1st} and \textsf{\small relRTCmpl}, the table also covers \emph{relative 50\% response time} (\emph{\textsf{\small relRT50}}); that is, the fraction of the overall execution time after which 50\% of all solutions of the corresponding query result have been computed.
\begin{table}[t]
\scriptsize
\centering
\begin{tabular}{|c||r|r||r|r||r|r|} \hline
	\cellcolor{gray!50} &
	\multicolumn{2}{c||}{\cellcolor{gray!50}\textbf{\textsf{\scriptsize relRT1st}}} &
	\multicolumn{2}{c||}{\cellcolor{gray!50}\textbf{\textsf{\scriptsize relRT50}}} &
	\multicolumn{2}{c|}{\cellcolor{gray!50}\textbf{\textsf{\scriptsize relRTCmpl}}}
\\ 
	\cellcolor{gray!50}\textbf{approach} &
	\cellcolor{gray!50}\textbf{worse} &
	\cellcolor{gray!50}\textbf{better} &
	\cellcolor{gray!50}\textbf{worse} &
	\cellcolor{gray!50}\textbf{better} &
	\cellcolor{gray!50}\textbf{worse} &
	\cellcolor{gray!50}\textbf{better}
\\ \hline
\TechReportVersion{
\DFXX	&	23.2	\!\% &	26.1	\!\% &	58.9	\!\% &	17.8	\!\% &	53.6	\!\% &	10.1	\!\% \\
\randomXX	&	13.0	\!\% &	27.5	\!\% &	58.9	\!\% &	8.2	\!\% &	59.4	\!\% &	8.7	\!\% \\
}
\PaperVersion{
}
\indegreeXX	&	21.7	\!\% &	21.7	\!\% &	65.8	\!\% &	4.1	\!\% &	50.7	\!\% &	5.8	\!\% \\ \hline
\rccOneXX	&	0.0	\!\% &	0.0	\!\% &	4.1	\!\% &	1.4	\!\% &	7.2	\!\% &	24.6	\!\% \\
\rccTwoXX	&	0.0	\!\% &	0.0	\!\% &	2.7	\!\% &	2.7	\!\% &	4.1	\!\% &	20.3	\!\% \\
\relOneXX	&	0.0	\!\% &	0.0	\!\% &	5.5	\!\% &	1.4	\!\% &	11.6	\!\% &	29.0	\!\% \\
\relTwoXX	&	0.0	\!\% &	0.0	\!\% &	11.0	\!\% &	0.0	\!\% &	2.9	\!\% &	26.1	\!\% \\ \hline
\isXX	&	7.2	\!\% &	31.9	\!\% &	15.1	\!\% &	27.4	\!\% &	26.1	\!\% &	10.1	\!\% \\
\PaperVersion{\hline}
\TechReportVersion{%
\isDecrXX	&	4.3	\!\% &	30.4	\!\% &	8.2	\!\% &	26.0	\!\% &	24.6	\!\% &	11.6	\!\% \\ \hline
}
\isrccOneXX	&	2.9	\!\% &	30.4	\!\% &	5.5	\!\% &	26.0	\!\% &	14.5	\!\% &	18.8	\!\% \\
\isrccTwoXX	&	5.8	\!\% &	33.3	\!\% &	5.5	\!\% &	24.7	\!\% &	13.0	\!\% &	26.1	\!\% \\
\isrelOneXX	&	0.0	\!\% &	33.3	\!\% &	2.7	\!\% &	24.7	\!\% &	15.9	\!\% &	26.1	\!\% \\
\isrelTwoXX	&	2.9	\!\% &	31.9	\!\% &	4.1	\!\% &	23.3	\!\% &	11.6	\!\% &	26.1	\!\% \\ \hline
\oracleXX	&	0.0	\!\% &	35.3	\!\% &	0.0	\!\% &	41.2	\!\% &	0.0	\!\% &	64.7	\!\% \\
 \hline
\end{tabular}
\caption{Percentage of cases in which the approaches
	achieve response times that are at least 10\% worse/better than \baseline.}
\label{tab:Comparison}
\end{table}

\TechReportVersion{
\begin{table*}[t]
\centering
	\subtable[]{%
		\scriptsize
\centering
\begin{tabular}{|c||r|r||r|r||r|r|} \hline
	\cellcolor{gray!50} &
	\multicolumn{2}{c||}{\cellcolor{gray!50}\textbf{\textsf{\scriptsize relRT1st}}} &
	\multicolumn{2}{c||}{\cellcolor{gray!50}\textbf{\textsf{\scriptsize relRT50}}} &
	\multicolumn{2}{c|}{\cellcolor{gray!50}\textbf{\textsf{\scriptsize relRTCmpl}}}
\\ 
	\cellcolor{gray!50}\textbf{approach} &
	\cellcolor{gray!50}\textbf{worse} &
	\cellcolor{gray!50}\textbf{better} &
	\cellcolor{gray!50}\textbf{worse} &
	\cellcolor{gray!50}\textbf{better} &
	\cellcolor{gray!50}\textbf{worse} &
	\cellcolor{gray!50}\textbf{better}
\\ \hline
%
\DFXX	&	28.6	\!\% &	22.9	\!\% &	74.3	\!\% &	20.0	\!\% &	65.7	\!\% &	8.6	\!\% \\
\randomXX	&	14.3	\!\% &	22.9	\!\% &	80.0	\!\% &	2.9	\!\% &	77.1	\!\% &	0.0	\!\% \\
\indegreeXX	&	28.6	\!\% &	20.0	\!\% &	91.4	\!\% &	0.0	\!\% &	62.9	\!\% &	2.9	\!\% \\ \hline
\rccOneXX	&	0.0	\!\% &	0.0	\!\% &	2.9	\!\% &	2.9	\!\% &	5.7	\!\% &	20.0	\!\% \\
\rccTwoXX	&	0.0	\!\% &	0.0	\!\% &	2.9	\!\% &	2.9	\!\% &	2.9	\!\% &	17.1	\!\% \\
\relOneXX	&	0.0	\!\% &	0.0	\!\% &	5.7	\!\% &	2.9	\!\% &	14.3	\!\% &	28.6	\!\% \\
\relTwoXX	&	0.0	\!\% &	0.0	\!\% &	14.3	\!\% &	0.0	\!\% &	2.9	\!\% &	22.9	\!\% \\ \hline
\isXX	&	5.7	\!\% &	31.4	\!\% &	20.0	\!\% &	31.4	\!\% &	25.7	\!\% &	17.1	\!\% \\
\isDecrXX	&	2.9	\!\% &	28.6	\!\% &	2.9	\!\% &	28.6	\!\% &	22.9	\!\% &	17.1	\!\% \\ \hline
\isrccOneXX	&	5.7	\!\% &	28.6	\!\% &	8.6	\!\% &	25.7	\!\% &	20.0	\!\% &	17.1	\!\% \\
\isrccTwoXX	&	5.7	\!\% &	31.4	\!\% &	8.6	\!\% &	28.6	\!\% &	14.3	\!\% &	25.7	\!\% \\
\isrelOneXX	&	0.0	\!\% &	31.4	\!\% &	2.9	\!\% &	28.6	\!\% &	17.1	\!\% &	20.0	\!\% \\
\isrelTwoXX	&	2.9	\!\% &	31.4	\!\% &	5.7	\!\% &	31.4	\!\% &	14.3	\!\% &	31.4	\!\% \\ \hline
\oracleXX	&	0.0	\!\% &	33.3	\!\% &	0.0	\!\% &	33.3	\!\% &	0.0	\!\% &	50.0	\!\% \\
 \hline
\end{tabular}
		\label{tab:ComparisonDense}
	}
\quad
	\subtable[]{%
		\scriptsize
\centering
\begin{tabular}{|c||r|r||r|r||r|r|} \hline
	\cellcolor{gray!50} &
	\multicolumn{2}{c||}{\cellcolor{gray!50}\textbf{\textsf{\scriptsize relRT1st}}} &
	\multicolumn{2}{c||}{\cellcolor{gray!50}\textbf{\textsf{\scriptsize relRT50}}} &
	\multicolumn{2}{c|}{\cellcolor{gray!50}\textbf{\textsf{\scriptsize relRTCmpl}}}
\\ 
	\cellcolor{gray!50}\textbf{approach} &
	\cellcolor{gray!50}\textbf{worse} &
	\cellcolor{gray!50}\textbf{better} &
	\cellcolor{gray!50}\textbf{worse} &
	\cellcolor{gray!50}\textbf{better} &
	\cellcolor{gray!50}\textbf{worse} &
	\cellcolor{gray!50}\textbf{better}
\\ \hline
\DFXX	&	17.6	\!\% &	29.4	\!\% &	44.7	\!\% &	15.8	\!\% &	41.2	\!\% &	11.8	\!\% \\
\randomXX	&	11.8	\!\% &	32.4	\!\% &	39.5	\!\% &	13.2	\!\% &	41.2	\!\% &	17.6	\!\% \\
\indegreeXX	&	14.7	\!\% &	23.5	\!\% &	42.1	\!\% &	7.9	\!\% &	38.2	\!\% &	8.8	\!\% \\ \hline
\rccOneXX	&	0.0	\!\% &	0.0	\!\% &	5.3	\!\% &	0.0	\!\% &	8.8	\!\% &	29.4	\!\% \\
\rccTwoXX	&	0.0	\!\% &	0.0	\!\% &	2.6	\!\% &	2.6	\!\% &	5.1	\!\% &	23.1	\!\% \\
\relOneXX	&	0.0	\!\% &	0.0	\!\% &	5.3	\!\% &	0.0	\!\% &	8.8	\!\% &	29.4	\!\% \\
\relTwoXX	&	0.0	\!\% &	0.0	\!\% &	7.9	\!\% &	0.0	\!\% &	2.9	\!\% &	29.4	\!\% \\ \hline
\isXX	&	8.8	\!\% &	32.4	\!\% &	10.5	\!\% &	23.7	\!\% &	26.5	\!\% &	2.9	\!\% \\
\isDecrXX	&	5.9	\!\% &	32.4	\!\% &	13.2	\!\% &	23.7	\!\% &	26.5	\!\% &	5.9	\!\% \\ \hline
\isrccOneXX	&	0.0	\!\% &	32.4	\!\% &	2.6	\!\% &	26.3	\!\% &	8.8	\!\% &	20.6	\!\% \\
\isrccTwoXX	&	5.9	\!\% &	35.3	\!\% &	2.6	\!\% &	21.1	\!\% &	11.8	\!\% &	26.5	\!\% \\
\isrelOneXX	&	0.0	\!\% &	35.3	\!\% &	2.6	\!\% &	21.1	\!\% &	14.7	\!\% &	32.4	\!\% \\
\isrelTwoXX	&	2.9	\!\% &	32.4	\!\% &	2.6	\!\% &	15.8	\!\% &	8.8	\!\% &	20.6	\!\% \\ \hline
\oracleXX	&	0.0	\!\% &	40.0	\!\% &	0.0	\!\% &	60.0	\!\% &	0.0	\!\% &	100.0	\!\% \\
 \hline
\end{tabular}
		\label{tab:ComparisonSparse}
	}
\caption{Percentage of cases in which the approaches achieve response times that are at least 10\% worse/better than the \baseline\ approach---focusing only on cases in test Webs that are (a) densely populated with bidirectional data links~(i.e., $\phi_1 \geq 0.66$), or (b)~sparsely populated with bidirectional data links~(i.e., $\phi_1 \leq 0.33$).}
\end{table*}
}

	\PaperVersion{For both \textsf{\small relRT1st} and \textsf{\small relRT50}, }%
	\TechReportVersion{For both \textsf{\small relRT1st} and \textsf{\small relRT50}, }%
we observe that \isrelOne\ is the best of the approaches tested (ignoring the \oracle\ approach which cannot be used in practice; cf.~Section~\ref{SecInText:OracleApproach}). Although the other intermediate solution driven approaches%
	\removable{~(\is, \TechReportVersion{\isDecr, }\isrelTwo, \isrccOne, \isrccTwo) }%
	\TechReportVersion{, as well as \DF\ and \random, }%
have a similarly high number of cases in which they are at least 10~\!\% better than the baseline, these approaches have a higher number of cases in which they are at least 10~\!\% worse.
\PaperVersion{%
\removable{We also notice that, as discussed in Section~\ref{sssec:Evaluation:Results:SolAware}, for \textsf{\small relRT1st}, the so\-lu\-tion-aware graph-based approaches~(\rccOne, \rccTwo, \relOne, \relTwo) behave like the \baseline.}
}
\TechReportVersion{%
As discussed
		\removable{in Section~\ref{sssec:Evaluation:Results:SolAware}},
	the so\-lu\-tion-aware graph-based approaches
		\removable{(\rccOne, \rccTwo, \relOne, and \relTwo)}
	behave like the \baseline\ approach.
	For \textsf{\small relRT50}, \isrelOne\ would again be the best~choice.

}

For \textsf{\small relRTCmpl}, we observe some differences. The hybrid approaches
	\removable{(\isrccOne, \isrccTwo, \isrelOne, \isrelTwo)}
	\PaperVersion{%
			\removable{still have a comparably high number of cases in which they are at least 10~\!\% better than the baseline, but they also} have a significant number of noteworthy cases in which they are worse.
		\is\ has }%
	\TechReportVersion{still have a comparably high number of cases in which they are at least 10~\!\% better than the baseline, but they also have a significant number of noteworthy cases in which they are worse. \is\ and \isDecr\ have }%
an even higher number of such worse cases. In contrast, the so\-lu\-tion-aware graph-based approaches are more suitable, with \relTwo\ being the best choice.

\TechReportVersion{

	To also examine whether the (un)suitability of approaches is affected by the density of bidirectional data links in the queried Web we break down the numbers in Table~\ref{tab:Comparison} into cases in test Webs that contain few or no bidirectional links~(i.e., $\phi_1 \leq 0.33$) versus test Webs that contain more bidirectional links~(i.e., $\phi_1 \geq 0.66$). Table~\ref{tab:ComparisonDense} focuses on the latter subset of cases, and Table~\ref{tab:ComparisonSparse} on the former. By comparing the numbers in these tables we make two general observations: First, the bidirectional interlinkage density indeed has an influence on the suitability of some of the approaches. However, second, the general trends remain the same.

\todo{Discuss notable differences in more detail!}
}

In summary, to return some solutions of query results as early as possible, \isrelOne\ appears to be the
	most suitable
choice among the approaches studied in this paper. However, if the objective is to reduce com\-plete-re\-sult response times, the so\-lu\-tion-aware graph-based approaches are usually better suited; in particular, \relTwo.

\PaperVersion{%
	In the extended version of the paper we additionally show that, by and large, these general findings are independent of whether the queried Web
		is densely populated with bidirectional data links~(i.e., $\phi_1 \geq 0.66$) or sparse~(i.e., $\phi_1 \leq 0.33$)~\cite{ExtendedVersion}.
}

\section{Conclusions} \label{sec:Conclusions}
This is the first paper that studies the problem of optimizing the response time of tra\-vers\-al-based query executions over Linked Data on the WWW.
	\PaperVersion{We show that }%
	\TechReportVersion{We first show that typical query optimization techniques for the process of constructing query results are unsuitable in this context, and, instead, }%
the fundamental problem is to fetch re\-sult-rel\-e\-vant data as early as possible. What makes this problem challenging is that data and data sources required for computing the query result are discovered recursively during query execution~(instead of being given a priori). Consequently, our optimization approach
	in this paper
is based on heu\-ris\-tics-based techniques to prioritize URI lookups during data retrieval. We study a diverse set of such techniques extensively; some of them can achieve
	\removable{noteworthy}
improvements over the baseline in a significant number of cases.

However, even for the best URI prioritization approaches in this paper, there exist
cases in which the \baseline\ approach achieves better response times. Moreover, a comparison to the \oracle\ approach shows that there is further
	room for improvement. A promising direction of future work
are approaches that collect statistics
	during (tra\-vers\-al-based) query executions
and leverage these statistics to optimize the response times for subsequent queries.

\balance
\bibliographystyle{abbrv}
\bibliography{main}

\TechReportVersion{%
\newpage
\onecolumn
\appendix

\section{FedBench Query Patterns}
\label{Appendix:FedBenchQueries}
\noindent
Some of the original FedBench Linked Data query patterns  are outdated (namely, $\mathsf{LD}_6$ to $\mathsf{LD}_{10}$). For our experiments, we slightly adjusted these patterns without changing their intent or their structural properties. We denote the original patterns by $\mathsf{LD}_6'$ to $\mathsf{LD}_{10}'$. In this appendix we list both versions of each of these patterns (i.e., the adjusted version that we have used for our experiments and the original version). For this list of patterns we assume the following prefix definitions:
\begin{footnotesize}
\begin{verbatim}
PREFIX rdf:  <http://www.w3.org/1999/02/22-rdf-syntax-ns#>
PREFIX rdfs: <http://www.w3.org/2000/01/rdf-schema#>
PREFIX owl:  <http://www.w3.org/2002/07/owl#>
PREFIX dct:  <http://purl.org/dc/terms/>
PREFIX dbowl:    <http://dbpedia.org/ontology/>
PREFIX dbprop:   <http://dbpedia.org/property/>
PREFIX drugbank: <http://www4.wiwiss.fu-berlin.de/drugbank/resource/drugbank/>
PREFIX foaf: <http://xmlns.com/foaf/0.1/>
PREFIX gn:   <http://www.geonames.org/ontology#>
PREFIX swc:  <http://data.semanticweb.org/ns/swc/ontology#>
PREFIX swrc: <http://swrc.ontoware.org/ontology#>
\end{verbatim}
\end{footnotesize}

\noindent
$\mathsf{LD}_2$:
\vspace{-1mm}
\begin{footnotesize}%
\begin{verbatim}
?proceedings swc:relatedToEvent <http://data.semanticweb.org/conference/eswc/2010> .
?paper swc:isPartOf ?proceedings .
?paper swrc:author ?p .
\end{verbatim}
\end{footnotesize}
\vspace{2mm}

\noindent
$\mathsf{LD}_5$:
\vspace{-1mm}
\begin{footnotesize}%
\begin{verbatim}
?a dbowl:artist <http://dbpedia.org/resource/Michael_Jackson> .
?a rdf:type dbowl:Album .
?a foaf:name ?n .
\end{verbatim}
\end{footnotesize}
\vspace{2mm}

\noindent
$\mathsf{LD}_9$ (adjusted):
\vspace{-1mm}
\begin{footnotesize}%
\begin{verbatim}
?x dct:subject <http://dbpedia.org/resource/Category:FIFA_World_Cup-winning_countries> .
?p dbprop:managerclubs ?x .
?p foaf:name "Luiz Felipe Scolari"@en .
\end{verbatim}
\end{footnotesize}
\vspace{2mm}

\noindent
$\mathsf{LD}_9'$ (original, outdated):
\vspace{-1mm}
\begin{footnotesize}%
\begin{verbatim}
?x skos:subject <http://dbpedia.org/resource/Category:FIFA_World_Cup-winning_countries> .
?p dbowl:managerClub ?x .
?p foaf:name "Luiz Felipe Scolari" .
\end{verbatim}
\end{footnotesize}
\vspace{2mm}

\noindent
$\mathsf{LD}_{10}$ (adjusted):
\vspace{-1mm}
\begin{footnotesize}%
\begin{verbatim}
?n dct:subject <http://dbpedia.org/resource/Category:Chancellors_of_Germany> .
?p2 owl:sameAs ?n .
?p2 <http://data.nytimes.com/elements/latest_use> ?u .
\end{verbatim}
\end{footnotesize}
\vspace{2mm}

\noindent
$\mathsf{LD}_{10}'$ (original, outdated):
\vspace{-1mm}
\begin{footnotesize}%
\begin{verbatim}
?n skos:subject <http://dbpedia.org/resource/Category:Chancellors_of_Germany> .
?n owl:sameAs ?p2 .
?p2 <http://data.nytimes.com/elements/latest_use> ?u .
\end{verbatim}
\end{footnotesize}
\section{Additional Test Queries}
\label{Appendix:SimulationQueries}
\noindent
This appendix lists the six conjunctive Linked Data queries that we have used for our synthetic test Webs. For this list we assume the following prefix definitions:
\begin{footnotesize}
\begin{verbatim}
PREFIX rdf:  <http://www.w3.org/1999/02/22-rdf-syntax-ns#>
PREFIX rdfs: <http://www.w3.org/2000/01/rdf-schema#>
PREFIX dc:   <http://purl.org/dc/elements/1.1/>
PREFIX rev:  <http://purl.org/stuff/rev#>
PREFIX bsbm: <http://www4.wiwiss.fu-berlin.de/bizer/bsbm/v01/vocabulary/>
PREFIX inP2: <http://www4.wiwiss.fu-berlin.de/bizer/bsbm/v01/instances/dataFromProducer2/>
PREFIX inP3: <http://www4.wiwiss.fu-berlin.de/bizer/bsbm/v01/instances/dataFromProducer3/>
PREFIX inP5: <http://www4.wiwiss.fu-berlin.de/bizer/bsbm/v01/instances/dataFromProducer5/>
PREFIX inR1: <http://www4.wiwiss.fu-berlin.de/bizer/bsbm/v01/instances/dataFromRatingSite1/>
PREFIX inV1: <http://www4.wiwiss.fu-berlin.de/bizer/bsbm/v01/instances/dataFromVendor1/>
\end{verbatim}
\end{footnotesize}

\newpage 

\noindent
\textbf{Q1} \\[1mm]
Query semantics: $\cMatch$-bag-semantics \\[1mm]
Query pattern: ~
\begin{minipage}[t]{10cm}
\begin{footnotesize}%
\begin{verbatim}
?offer bsbm:vendor inV1:Vendor1 .
?offer bsbm:product ?product .
?product bsbm:producer inP2:Producer2 .
\end{verbatim}
\end{footnotesize}
\end{minipage} \\[2mm]
Seed URIs: ~\texttt{inV1:Vendor1}, \texttt{inP2:Producer2}
\vspace{4mm}

\noindent
\textbf{Q2} \\[1mm]
Query semantics: $\cMatch$-bag-semantics \\[1mm]
Query pattern: ~
\begin{minipage}[t]{10cm}
\begin{footnotesize}%
\begin{verbatim}
inP5:Product184 bsbm:producer ?producer .
inP5:Product184 rdf:type ?type .
?product2 rdf:type ?type .
?product2 bsbm:producer ?producer .
\end{verbatim}
\end{footnotesize}
\end{minipage} \\[2mm]
Seed URI: ~\texttt{inP5:Product184}
\vspace{4mm}

\noindent
\textbf{Q3} \\[1mm]
Query semantics: $\cMatch$-bag-semantics \\[1mm]
Query pattern: ~
\begin{minipage}[t]{10cm}
\begin{footnotesize}%
\begin{verbatim}
?review bsbm:reviewFor inP3:Product128 .
?review bsbm:rating1 10 .
?review rev:reviewer ?reviewer .
?reviewer bsbm:country ?country .
\end{verbatim}
\end{footnotesize}
\end{minipage} \\[2mm]
Seed URI: ~\texttt{inP3:Product128}
\vspace{4mm}

\noindent
\textbf{Q4} \\[1mm]
Query semantics: $\cMatch$-bag-semantics \\[1mm]
Query pattern: ~
\begin{minipage}[t]{10cm}
\begin{footnotesize}%
\begin{verbatim}
inP3:Product128 bsbm:producer       ?producer .
inP3:Product128 bsbm:productFeature ?feature .
?product2 bsbm:productFeature ?feature .
?product2 bsbm:producer       ?producer .
\end{verbatim}
\end{footnotesize}
\end{minipage} \\[2mm]
Seed URI: ~\texttt{inP3:Product128}
\vspace{4mm}

\noindent
\textbf{Q5} \\[1mm]
Query semantics: $\cMatch$-bag-semantics \\[1mm]
Query pattern: ~
\begin{minipage}[t]{10cm}
\begin{footnotesize}%
\begin{verbatim}
inR1:Review110 bsbm:reviewFor ?product .
?product bsbm:productFeature ?feature .
?feature rdfs:label ?featureLabel .
\end{verbatim}
\end{footnotesize}
\end{minipage} \\[2mm]
Seed URI: ~\texttt{inR1:Review110}
\vspace{4mm}

\noindent
\textbf{Q6} \\[1mm]
Query semantics: $\cMatch$-bag-semantics \\[1mm]
Query pattern: ~
\begin{minipage}[t]{10cm}
\begin{footnotesize}%
\begin{verbatim}
?review bsbm:reviewFor inP3:Product128 .
?review bsbm:rating1 ?rating .
?review dc:title ?reviewTitle .
\end{verbatim}
\end{footnotesize}
\end{minipage} \\[2mm]
Seed URI: ~\texttt{inP3:Product128}
\vspace{4mm}

}

\end{document}